\documentclass[sigconf,authorversion]{acmart}
\AtBeginDocument{%
  \providecommand\BibTeX{{%
    \normalfont B\kern-0.5em{\scshape i\kern-0.25em b}\kern-0.8em\TeX}}}

\setcopyright{none}
\copyrightyear{2023}
\acmYear{2023}
\setcopyright{acmlicensed}
\acmConference[RAID '23]{The 26th International Symposium on Research in Attacks, Intrusions and Defenses}{October 16--18, 2023}{Hong Kong, Hong Kong}
\acmBooktitle{The 26th International Symposium on Research in Attacks, Intrusions and Defenses (RAID '23), October 16--18, 2023, Hong Kong, Hong Kong}

\acmDOI{10.1145/3607199.3607208}

\usepackage{xcolor}

\usepackage{amsfonts}
\usepackage{multirow}
\usepackage{float}
\usepackage{soul}

\usepackage{algorithmic}  
\usepackage[ruled,vlined,linesnumbered]{algorithm2e}

\usepackage{breakurl}           % break too-long urls in refs
\usepackage{url}                % allow \url in bibtex for clickable links
\usepackage[]{hyperref}         % ...clickable refs within pdf...
\hypersetup{                    % ...like so
  colorlinks,
  linkcolor={green!80!black},
  citecolor={red!70!black},
  urlcolor={blue!70!black}
}

\usepackage[inline]{enumitem}

\begin{document}

\title{Looking Beyond IoCs: Automatically Extracting Attack Patterns from External CTI}

\author{Md Tanvirul Alam}
\affiliation{%
  \institution{Rochester Institute of Technology}
  \city{Rochester, New York}
  \country{USA}}
\email{tanvirul.alam@mail.rit.edu}

\author{Dipkamal Bhusal}
\affiliation{%
  \institution{Rochester Institute of Technology}
  \city{Rochester, New York}
  \country{USA}}
\email{db1702@rit.edu}

\author{Youngja Park}
\affiliation{%
  \institution{IBM Research}
  \city{Yorktown Heights, New York}
  \country{USA}}
\email{young\_park@us.ibm.com}

\author{Nidhi Rastogi}
\affiliation{%
  \institution{Rochester Institute of Technology}
  \city{Rochester, New York}
  \country{USA}}
\email{nxrvse@rit.edu}


\begin{abstract}
Public and commercial organizations extensively share cyberthreat intelligence (CTI) to prepare systems to defend against existing and emerging cyberattacks. However, traditional CTI has primarily focused on tracking known threat indicators such as IP addresses and domain names, which may not provide long-term value in defending against evolving attacks. To address this challenge, we propose to use more robust threat intelligence signals called attack patterns. \textsf{LADDER} is a knowledge extraction framework that can extract text-based attack patterns from CTI reports at scale. The framework characterizes attack patterns by capturing the phases of an attack in Android and enterprise networks and systematically maps them to the MITRE ATT\&CK pattern framework. \textsf{LADDER} can be used by security analysts to determine the presence of attack vectors related to existing and emerging threats, enabling them to prepare defenses proactively. We also present several use cases to demonstrate the application of \textsf{LADDER} in real-world scenarios. Finally, we provide a new, open-access benchmark malware dataset to train future cyberthreat intelligence models. 
\end{abstract}






\maketitle

\section{Introduction}

% What is CTI and why is it Important with examples.
Cyber Threat Intelligence (CTI) offers crucial insights into the rapidly evolving cyber threat landscape. This information includes any evidence to identify and assess the associated threats, such as indicators of compromise (IOCs), IP addresses, domain names, and file hashes, and any associated tactics, techniques, and procedures (TTPs) used by the attacker(s). For instance, CTI can provide comprehensive, contextual information on emerging threats like the advanced persistent threat (APT), ScarCruft \cite{singh_selvan_2023}. Also known as APT37, the cyber threat intelligence on ScarCruft reported that the APT targets ``individuals in South Korean organizations'' with the primary objective of ``cyber espionage.'' The CTI reports that the APT achieves this through ``data exfiltration of selected file formats'' and uses MD5 hashes, known as IoCs. Therefore, organizations need to leverage CTI to understand adversary tactics and goals, prevent future attacks, and shorten the time for remedial measures. Security analysts aggregate, clean, analyze, evaluate, and contextualize cyberattack information to produce comprehensive reports. The goal is to enhance cybersecurity-related decision-making for organizations facing similar threats \cite{nist_cti}. Later, the CTI report can be disseminated to interested parties through paid subscriptions or free resources such as blogs, bulletins, news, and reports \cite{bouwman2020different}. However, once the CTI is received, it can still present significant challenges for organizations that want to make this information actionable. 
Some of the key challenges are:
\begin{enumerate}
    \item \textbf{Timely identifying and extracting pertinent information and integrating threat signals into internal defense infrastructure.} While traditional forms of intelligence collected from static indicators such as malware hash and IP address can be obtained through pattern matching, they fail to detect new generations of cyber threats since attackers can modify them to evade detection \cite{tounsi2018survey}. For example, changing a single bit in the binary can alter the malware hash. As a result, there is a growing demand for more tactical threat intelligence extraction from CTI sources that are more robust to evolving adversary evasion TTPs. 
    
    % Thus, automating the extraction and integration of CTI IoC is important.

    \item \textbf{Extracting pertinent information from CTI sources due to the unstructured, semi-structured format and presence of noisy information.} While some research focuses on automating the detection and extraction of Indicators of Compromise (IoC)  \cite{husari2017ttpdrill}, tracking IoCs is not the same as threat hunting as there is little overlap of shared IoCs among organizations, and potential delay in IoCs usage and detection \cite{bouwman2020different}. To address these issues, researchers have focused on extracting attack patterns from CTI \cite{husari2017ttpdrill,li2022attackg}.

    \item \textbf{ Extracting TTPs due to their natural language descriptions and the evolution of various sub-techniques over time.} Mapping extracted attack patterns to a standard format can decrease redundancy during further analysis. However, this is often difficult due to the ambiguity inherent in the text. Rule-based systems for attack pattern extraction may be inadequate when an attack pattern (or TTP) lacks associated trigger words. Consequently, a machine-learning-based approach is required to extract higher-level tactical intelligence from unstructured CTI texts.

\end{enumerate}

 %Additionally, utilizing a large volume of CTI can result in information redundancy. Some works have proposed aggregating information in a cybersecurity knowledge graph to make it easier to analyze and utilize the gathered information for different applications \cite{Piplai2020CreatingCK,Kiesling2019TheSK,Sarhan2021OpenCyKGAO,satvat2021extractor}.

\par

\begin{figure*}[ht]
\centering
\includegraphics[width=1\textwidth,height=.28\textwidth]{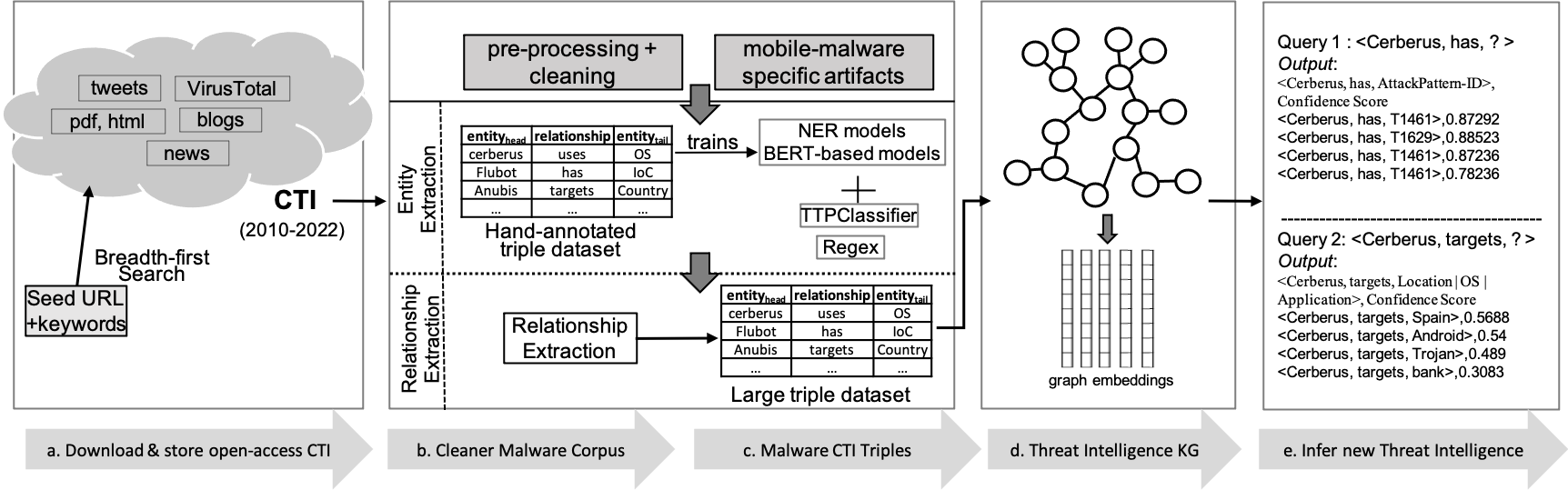}
\caption{Proposed framework: \textsf{LADDER} and the five component modules. (a) Extracts CTI using crawlers, (b) Pre-processes and prepares for entity and relationship extraction, (c) generates data in the form of triples, (d) creates a knowledge graph by combining the triples and uses graph embedding methods to pack every triple's properties into a vector with smaller dimensions, and (e) returns instances when to queried by the analyst. (a)-(d) are training step, (e) is inference step.}
\label{fig:ladder}
\end{figure*}

We present \textsf{LADDER}, a framework to automatically extract Tactics, Techniques, and Procedures (TTPs) and other relevant information from CTI sources related to the malware and APTs. We restructure this information using an ontology \cite{christian2021ontology} and \textsf{TTPClassifier} into a knowledge graph (KG) to enable predictive analysis (see Figure \ref{fig:ladder}). \textsf{TTPClassifier} utilizes a novel machine-learning algorithm for TTP extraction from CTI reports (includes IoCs and TTPs). It categorizes the TTPs into standardized MITRE ATT\&CK~\cite{mitre_attck} pattern IDs, as shown in Section \ref{ttpclassifier}. The \textsf{TTPClassifier} enables analysts to learn and analyze attack campaigns for existing or emerging threats, ultimately helping to preempt potential attacks on their organizations. Our proposed framework addresses a critical gap in the automated extraction and analysis of CTI, providing a valuable tool for organizations to enhance their threat detection capabilities. The main contributions are:

% \begin{enumerate}

% \item We propose \textsf{LADDER}, a threat intelligence aggregation and analysis framework that automatically extracts and restructures attack patterns and IoCs as evidence of attacks found in diverse, unstructured CTI. \textsf{LADDER} also classifies them according to ATT\&CK pattern techniques described by MITRE.

% \item We demonstrate the effectiveness of \textsf{LADDER} and its security applications for SOC analysts by accurately extracting attack patterns, performing predictive analysis of malware behavior, threat conducting threat hunting, and attributing APT groups. 

% \item We provide a new, open-access benchmark malware dataset to train future cyberthreat intelligence models. It consists of 140,447 tokens, including manually annotated 11,555 named entities and 5,499 relations\cite{anon}.
% \end{enumerate}

\begin{enumerate}

\item We propose \textsf{LADDER}, a threat intelligence aggregation and analysis framework that automatically extracts and restructures attack patterns and IoCs as evidence of attacks found in diverse, unstructured CTI. \textsf{LADDER} also classifies them according to ATT\&CK pattern techniques described by MITRE. This is the first work to include standardized ATT\&CK patterns in the KG with other forms of threat intelligence.

\item We demonstrate the effectiveness of \textsf{LADDER} and its security applications for security analysts by accurately extracting attack patterns, performing predictive analysis of malware behavior, threat conducting threat hunting, and attributing APT groups. To the best of our knowledge, this is the first work to utilize KG for malware attack pattern prediction. 

\item We provide a new, open-access benchmark malware dataset to train future cyberthreat intelligence models. It consists of 140,447 tokens, including manually annotated 11,555 named entities and 5,499 relations. The dataset and code are publicly available.\footnote{\url{https://github.com/aiforsec/LADDER}}
\end{enumerate}
\section{Motivating Example}\label{motivatingExample}

\begin{figure*}[ht]
\centering
\includegraphics[width=1\textwidth]{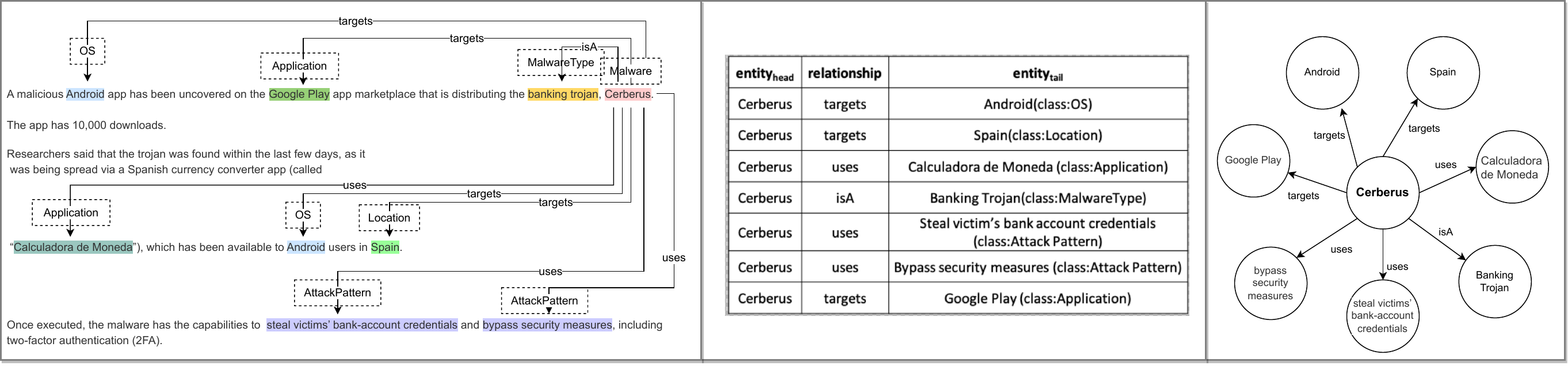}
\caption{For malware Cerberus, (a) Annotated CTI using BRAT, (b) Triples created from the annotated snippet, (c) Knowledge graph from the triples (best viewed when zoomed).}
\label{fig:Cerberus}
\end{figure*}

To inspire our research, We present an example of how \textsf{LADDER} can be utilized by security analysts to leverage CTI reports. We study the CTI of malware \textsf{Cerberus}, a trojan horse that targets Android mobile phone banking credentials. This CTI provides a comprehensive description of the malware's capabilities and a detailed account of its tactics, techniques, and procedures (TTPs), also called attack patterns. Furthermore, the CTI encompasses different vulnerabilities in various business technologies, including email, domains, and mobile devices. The following excerpt from a \textsf{Cerberus} CTI posted on ThreatPost \cite{cerberus-unleased} illustrates this:

\textit{"...A malicious Android app has been uncovered on the Google Play app marketplace that is distributing the banking Trojan, \textsf{Cerberus}. The app has 10,000 downloads. Researchers said that the trojan was found within the last few days, as it was being spread via a Spanish currency converter app (called "Calculadora de Moneda"), which has been available to Android users in Spain. Once executed, the malware has the capabilities to steal victims' bank-account credentials and bypass security measures, including two-factor authentication (2FA)..."}.

Figure \ref{fig:Cerberus} shows the transformation from a CTI report to a knowledge graph for ``\textsf{Cerberus}''. In Section \ref{sec:system}, we detail the process of entity extraction (e.g., application), triple generation $\langle$malware, targets, application$\rangle$, and knowledge graph construction (combination of all triples). The table \ref{fig:Cerberus}(b) shows entity classes and relationships following our ontology described in Section \ref{sec:concepts}. It \textsf{isA} \textit{``banking Trojan"}, \textsf{targets} \textit{``Spain, class:Location"}, \textit{``Android, class:OS"} devices,  and uses \textsf{attack patterns} such as \textit{``Bypass security measures"}, and \textit{``Steal victim's bank account credentials"}. 

The entity class definitions for \textsf{Malware, Attack Pattern, Location, OS, Application} (see Section \ref{sec:concepts}) map to existing threat intelligence ontology classes ~\cite{rastogi2020malont,christian2021ontology}. However, they have been adapted for CTI and security logs and follow the STIX2.1 framework for TTP and IoC exchange. Relationships between them are pre-defined within the same ontology. We extract triples from CTI reports to utilize historical malware information, including attack patterns, to train a threat intelligence model. 

For instance, in Section \ref{sec:threathunt}: use cases, we show how an analyst maps attack patterns based on external CTI to internal security logs. Section \ref{sec:usecase} shows how an analyst queries our knowledge graph using the attack patterns extracted from the CTI. The graph can ``infer'' potential attack patterns that the same malware might attempt, even if these patterns have not previously been reported or observed. Knowledge of the MITRE ATT\&CK is advantageous to the analyst as mitigation techniques are provided for each attack pattern, enabling even a less experienced security analyst to take timely actions. Using this knowledge, an analyst can take proactive measures to prevent or deter adversaries from causing damage to the internal network.

\section{Background}

\subsection{Cyber Threat Intelligence}
Cyber Threat Intelligence (CTI) is evidence-based knowledge about existing or emerging cyber threats, which can facilitate decision-making processes in response to cyberthreats. CTI should be relevant (related to an objective), actionable (prompts a response to a threat), and valuable (contributes to a business outcome) \cite{dalziel2014define}. CTI can be collected both internally within the organization and externally. Organizations can gather internal intelligence from the system and network endpoint logs. External threat intelligence is acquired (freely or at a cost) from sources outside the organization. CTI is collected from security bulletins, the dark web, hacking-related websites, public threat reports, source code repositories, and social media platforms like GitHub and Reddit \cite{tundis2022feature}.
\par
\textbf{Unstructured CTI. }Open-source CTI comes in both \textbf{structured} and \textbf{unstructured} forms. Many standards have developed over the years for structured CTI sharing-- STIX (Structured Threat Information eXpression) \cite{barnum2012standardizing}, TAXII (Trusted Automated eXchange of Indicator Information) \cite{connolly2014trusted}, ATT\&CK (AdversarialTactics, Techniques \& Common Knowledge) \cite{strom2018mitre} and many others. Structured CTIs allow efficient automation and collection of information from diverse sources of CTI. However, generating such structured threat reports is time-consuming and requires a lot of manual labor. As such, many public threat reports are provided in an unstructured format by different security farms such as Symantec \cite{symantec}, McAfee \cite{mcafee}, Kaspersky \cite{kaspersky}. Although these reports are more readily available, extracting relevant information for a specific organization or security analyst is still challenging. Without information extraction tools, security analysts may become overburdened with the large volume of information available \cite{liao2016acing}. This issue becomes even more significant when dealing with tactical threat intelligence, such as adversary tactics, techniques, and procedures. As such, there is a need for automatic information extraction from diverse unstructured CTI sources to make the information actionable.

\subsection{Collecting and Structuring CTI Concepts}\label{sec:concepts} 

\begin{figure}[t]
\centering
\includegraphics[width=.3\textwidth, height=.3\textwidth]{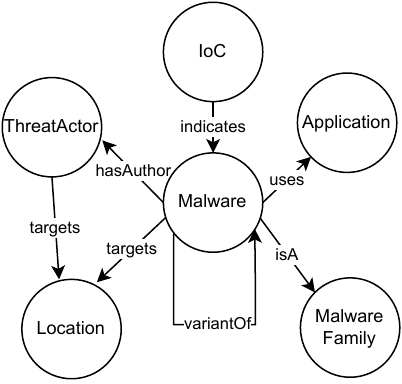}
\caption{Part of the ontology used in the study}
\label{fig:onto}
\end{figure}

We create knowledge graphs from CTI because these graphs can transform unstructured information about CTI into a structured format. They can also store a vast amount of domain-specific information in the form of triples representing pairs of entities and the relationship between them \cite{nickel2015review}. This approach necessitates capturing context from threat intelligence information and representing it in a structured format using RDF expressions, \textit{<subject, predicate, object>}. To facilitate this process, we adapt from existing ontologies on malware \cite{rastogi2020malont,christian2021ontology}. The cyber threat concepts in the knowledge graph include \textsf{Malware, Malware Type, Application, Operating System, Organization, Person, Time, Threat Actor, Location,} and \textsf{Attack Pattern}. These concepts are connected through ten relations, including \textsf{isA, targets, uses, hasAuthor, hasAlias, indicates, discoveredIn, exploits, variantOf,} and \textsf{has}. Figure \ref{fig:onto} shows a part of the ontology. Some triples may convey low-level threat intelligence, like the hash or IP addresses associated with the malware. Others can capture higher-level intelligence, such as the applications targeted by malware or their attack patterns. We provide details of cybersecurity concepts in the Appendix. By structuring the open-access CTI in this manner, we enable efficient analysis and automated information extraction for security analysts.

\subsection{Attack Patterns}
Attack patterns depict an attacker's methods to achieve a tactical goal offering a high-level insight into the motivation behind an attack. For instance, an adversary may encrypt files on a device to prevent access until a ransom is paid. MITRE's Adversarial Tactics, Techniques, and Common Knowledge (ATT\&CK) framework enumerates common attack patterns with 66 unique techniques for mobile platforms alone.\footnote{\url{https://attack.mitre.org/techniques/mobile/}} Unlike Indicators of Compromise (IoCs), which may be short-lived, Threat Tactic Techniques and Procedures (TTPs) deliver long-term intelligence for cyber threat analysis. TTPs can assist organizations in evaluating the effectiveness of their security measures against current threats and aid attackers' emulation for testing and validating defenses against prevalent techniques \cite{miller2018automated, applebaum2016intelligent}.

% In formal terms, a Knowledge Graph is defined as $\mathcal {G} = {\mathcal {E}, \mathcal {R}, \mathcal {F}}$, where $\mathcal{E}$, $\mathcal{R}$, and $\mathcal{F}$ are sets of entities, relations, and facts, respectively \cite{ji2021survey}. A fact is represented as an RDF triple $(h,r, t) \in \mathcal{F}$, where $h$ denotes the head entity, $t$ denotes the tail entity, and $r$ denotes the relation between them.
% \input{sections/approach}
\section{System Design}\label{sec:system}

% {\textcolor{red}{Move to related work}}
% We collected a large number of threat intelligence reports from public CTI sources to conduct our research. We sampled a subset of these reports to create a hand-annotated corpus for training, evaluating and verifying \textsf{LADDER}. We also downloaded reports from VirusTotal using the academic API to enhance the knowledge graph and analyse improvements in inference due to added knowledge.
\subsection{Dataset Collection}
To our knowledge, no public dataset containing triples extracted from CTI sources exists. To ensure the highest quality of triples for our knowledge graph creation, we curated CTI reports related to 36 malware, including \textit{Cerberus, Rotexy, Judy, Gooligan,} and \textit{SpyNote RAT} listed on the MITRE website; representative of Android malware during 2015-2022. Many attack patterns in these reports were paraphrased and mapped to the MITRE ATT\&CK framework. These reports are authored by security analysts from reputable organizations such as McAfee, Symantec, and Kaspersky and provide natural language descriptions of the malware emergence, propagation, attack patterns, and IoCs.

%vAnalyzing these reports is the same as described in the five modules of \textsf{LADDER} in Section \ref{sec:CTI}. However, the information extraction module is missing, as concepts were hand-annotated by security researchers, mainly comprising graduate students studying machine learning and security and overseen by their advisors. An example annotation is shown in Figure \ref{fig:Cerberus}.

% \begin{figure*}
% \begin{multicols}{2}
%     \fbox{\includegraphics[width=.45\textwidth, height=.21\textwidth]{figures/cerberusRAT.png}}\par 
%     \fbox{\includegraphics[width=.49\textwidth]{figures/CTIsnap.png}}\par 
%     \end{multicols}
% \caption{caption here}
% \label{fig:brat}
% \end{figure*}

% \begin{figure}[th]
% \centering
% \scriptsize
% \fbox{\includegraphics[width=.45\textwidth]{figures/Distributions.png}}
% \caption{Distributions in hand-annotated corpus (up-left) Entity, (up-right) Relationship (bottom) Number of attack patterns per malware %\textcolor{red}{UPDATE}
% }.
% \label{fig:atk-dist}
% \end{figure}

\begin{figure}[th]
\tiny
\centering
\resizebox{1\linewidth}{!}{
\includegraphics[width=.5\textwidth]{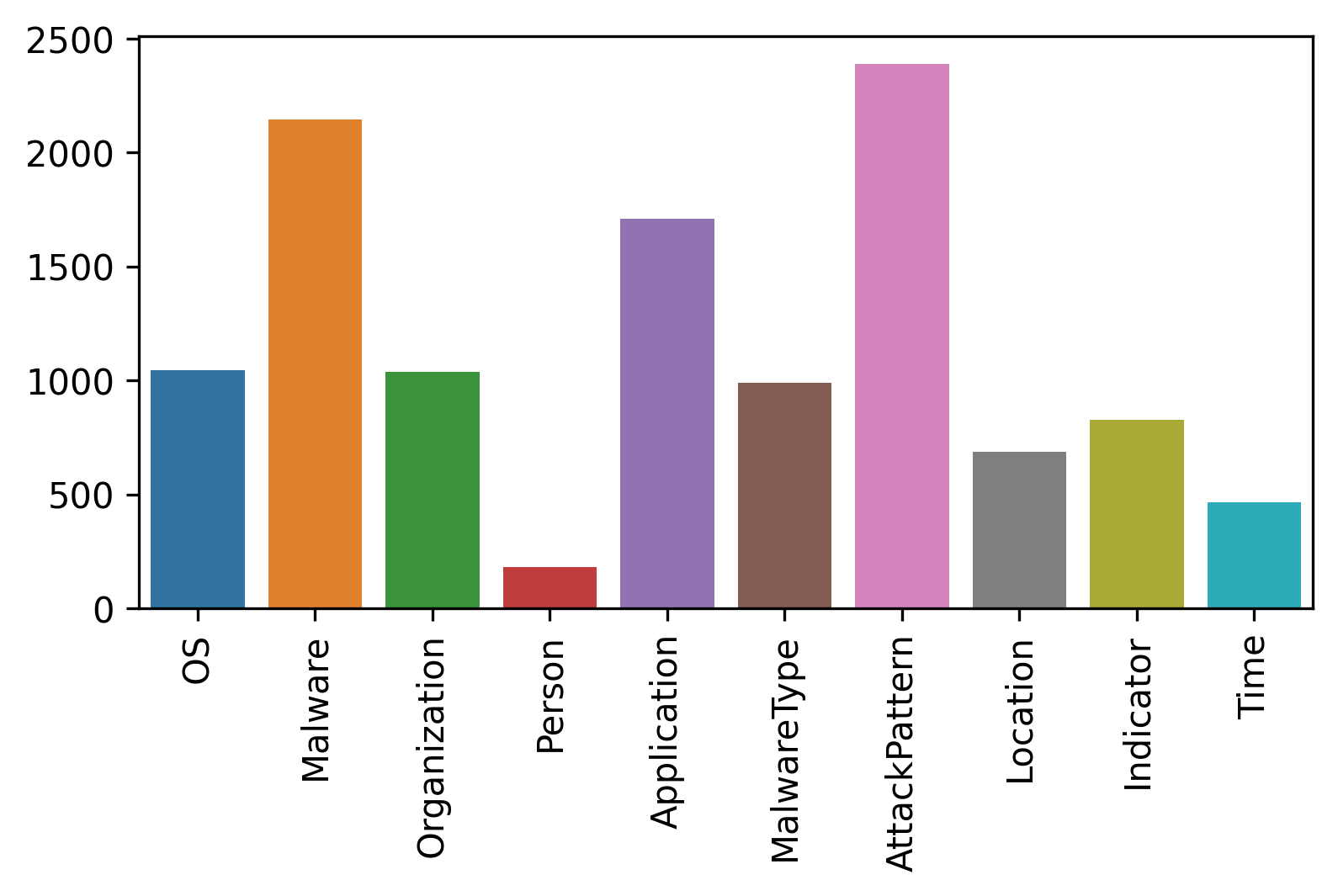}
\includegraphics[width=.5\textwidth]{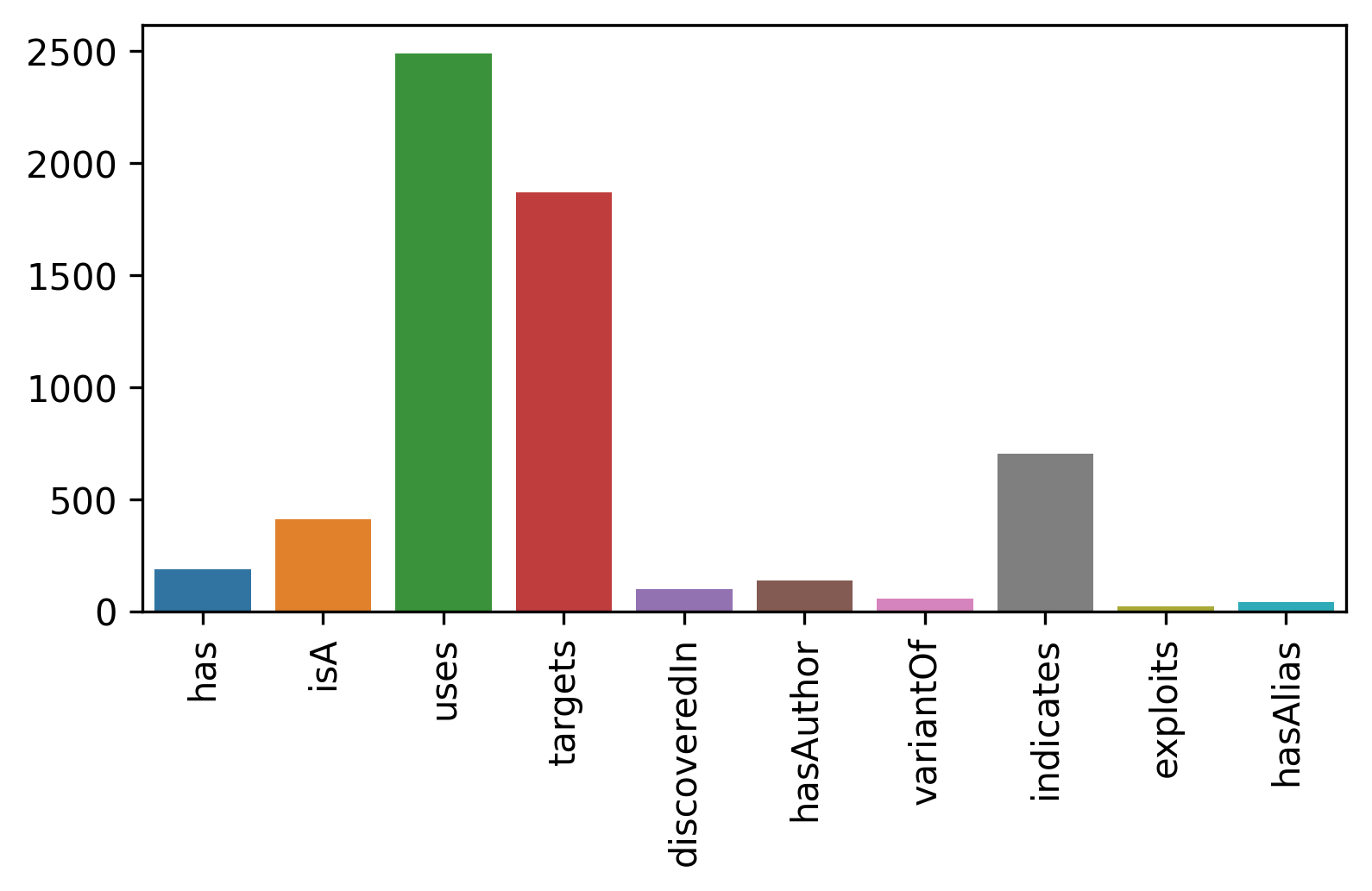}
}
\caption{Entity (l) and relationships (r) distribution.}
\label{fig:stat-ent}
\end{figure}

% \begin{figure}[th]
% \tiny
% \centering
% \resizebox{0.8\linewidth}{!}{%
% \includegraphics[width=.45\textwidth]{figures/relations.png}
% }
% \caption{Distributions for relations in the corpus}
% \label{fig:stat-rel}
% \end{figure}
\textbf{Annotation. }We employed the widely used open-source annotation tool,  BRAT \cite{Stenetorp2012bratAW} to manually annotate threat concepts and their relationships, as explained in Section \ref{sec:concepts}. An example of our annotation is shown in Figure \ref{fig:Cerberus}. It is important to note that some entities may classify into different classes depending on their context. For instance, the terms \textit{Facebook, Twitter} and \textit{Instagram} can be classified as either \textsf{Organization} or \textsf{Application} depending on the context. We also annotated attack patterns as concepts, but these often include other concepts within them due to their extended length. In such cases, we annotated the larger text and the smaller entities within the text as an attack pattern. For example, the attack pattern \textit{``break Android's application sandbox''} includes the annotation for \textit{Android} as \textit{OS}. We show the distribution of different entities in Figure \ref{fig:stat-ent} (l), with AttackPattern being the most common entity type, followed by Malware and Application. The distributions of relationships between entities are shown in Figure \ref{fig:stat-ent} (r). The relationship \textit{uses} has the highest count because it links attack patterns with the associated Malware. The \textit{Indicates} relation connects compromised indicators with the representative Malware. The \textit{targets} relationship links \textsf{Malware} with \textsf{ThreatActor} and other entities. Finally, the \textit{isA} relationship indicates a broader category or family of Malware.

% \textsf{Malware} and the \textsf{Application} or \textsf{Location} it targets. \textit{isA} draws a relationship between a broader category or family of the malware. If malware is developed based on another existing malware, they connect with the \textit{variantOf} relationship. \textit{hasAlias} shows different names of the same malware.
%We obtained an inter-annotator agreement of ..... Cohen's Kappa for a pair of annotators. (describe the result depending on what value was obtained)

% More details on the entity and relation classes can be found in the appendix. 

% Overall, our dataset contains 11,555 named entity tags and 5,499 relations among them.

\subsection{Information Extraction from CTI for Threat Intelligence Graph}\label{infoExtract}

\textbf{Dataset Crawler. }The annotated dataset enables us to train machine learning models for information extraction tasks. The trained models can then be used for knowledge graph extraction from a broader set of CTI reports. To this end, we developed a high-performance web crawler that scraped over 12,000 relevant unstructured open-access CTI reports from public URLs. We focused our search on security and technology companies and technology news reporting companies to ensure report quality. The crawler used a breadth-first search (BFS) starting from a seed URL belonging to a security or technical CTI website. It saved all URLs mentioned on the starting page and assessed their text for relevance. A relevant page included detailed descriptions of malware, such as the presence of malware-related keywords within the first $n$ words of the article (where $n<100$). The crawler saved the URL and its text if the page was relevant. Then, the text cleaner processed the content by removing HTML tags and images. We also extracted temporal information from the reports for longitudinal analysis using a heuristic-based approach. Specifically, we used the NER model provided in the Flair \cite{akbik2019flair} NLP library to extract the first \textit{DATE} entity found in the top five sentences, if present. We verified the date and extracted the year using the python \textsf{datefinder} \cite{datefinder} library. We limited our research to threat reports published between 2010-2021. Details of web crawling Algorithm \ref{algo:scraper} are in the Appendix.

\textbf{Entity Extraction. } Once the threat reports are pre-processed and cleaned, we extract different entity classes using state-of-the-art natural language processing techniques. We fine-tune Transformer-based \cite{vaswani2017attention} pretrained language model using our hand-annotated dataset for classes including \textsf{Malware, AttackPattern, Application, OS, Organization, Person, Time, Location}. We adopt this approach because such large-scale language models are highly effective in numerous downstream NLP tasks with limited labeled data \cite{devlin2018bert}. Specifically, we use three transformer variants in our experiments: BERT \cite{devlin2018bert}, RoBERTa\cite{liu2019roberta} and XLM-RoBERTa\cite{conneau2019unsupervised}. These models use powerful attention mechanisms to capture context and extract evolving security concepts like attack patterns and malware. While BERT and RoBERTa are pre-trained on English corpora, XLM-RoBERTa is multilingual. These models leverage their powerful attention mechanisms to capture the context and extract evolving security concepts such as attack patterns and malware. We use the \textit{tner} library \cite{ushio2022t} to fine-tune the models, which enables us to achieve high accuracy in entity recognition. Specifically, we take the hidden layer representation from the transformer model and add a classification layer with ten neurons corresponding to the nine entity classes and a special no entity (\textit{O}) token, illustrating that a token does not belong to any entity classes. On the other hand, we use pattern matching to extract IoCs such as URL, IP address, email, and file name. We describe the regular expressions used for pattern matching and their corresponding entity types in the Appendix.

\subsection{Attack Pattern Extraction} \label{ttpclassifier}

Attack pattern extraction and labeling pose unique challenges. Attack patterns comprise a larger block of text that describes a cyber threat action rather than a single named entity. Sometimes, an attack pattern may include other entity types within its description. For example, in the sentence "Cerberus is capable of generating an instance of TeamViewer on mobile," the attack pattern phrase "capable of generating an instance of TeamViewer on mobile" contains the entity "TeamViewer" of type "Application" within it. These nuances necessitate a different approach to extraction and labeling, considering the broader context of the attack pattern and its relationship to other entities in the text. To address these challenges, we present \textsf{TTPClassifier}, a novel approach for extracting attack patterns from threat reports. This approach comprises three sub-tasks: relevant sentence extraction, attack phrase identification \& extraction, and mapping attack patterns to the MITRE ATT\&CK. We discuss these sub-tasks below:
\begin{enumerate}
    \item \textbf{Relevant Sentence Extraction. }Firstly, we identify sentences that contain an attack description. We approach this task as a binary sentence classification problem, where sentences containing one or more attack patterns are labeled positive and the rest negative. During training, we use all annotated positive instances and randomly select an equal number of negative instances from the remaining sentences to form a balanced dataset. We fine-tune pre-trained transformer models for this classification task, adding a linear layer with two neurons. For RoBERTa models, we add a hidden layer before the linear layer.

    \item \textbf{Attack Pattern Identification \& Extraction. }Secondly, we identify the relevant parts of sentences containing attack pattern descriptions for those predicted as positive, i.e., having at least one attack pattern. We use a sequence tagging model for this subtask, similar to entity extraction. Using two classes, the model predicts whether each word in the sentence is part of an attack pattern description. Some sentences may contain \textit{more than one} attack pattern, necessitating the combination of each contiguous block tagged with the attack pattern entity into a single attack pattern description. Consider the following example:
\small
\textit{``The malware can covertly send and steal SMS codes, open tailored overlays for various online banks, and steal 2FA-codes".} 
\normalsize
This sentence contains three attack patterns: (a) \textit{``covertly send and steal SMS codes"}, (b) \textit{``open tailored overlays for various online banks"}, and (c) \textit{``steal 2FA-codes"}. Since \textsf{TTPClassifier} makes predictions for individual tokens, we combine each contiguous block tagged with the attack pattern entity into a single attack pattern description. During post-processing, we discard invalid extractions, e.g., those that do not contain verbs.

    \item \textbf{Mapping to ATT\&CK ID. }Finally, we map each extracted attack pattern to standardized ATT\&CK techniques. Although ATT\&CK has both techniques and sub-techniques, we only consider the former and map each extracted sequence to one technique. Due to the large number of potential classes and significant annotation effort required to match an attack pattern to its corresponding ATT\&CK ID, we adopt a semantic similarity-based approach for the mapping task. We first compute embeddings for the extracted attack pattern phrases using a pre-trained sentence transformer model \cite{reimers-2019-sentence-bert}. We use embeddings from the title and description of the ATT\&CK ID described on the website for improved learning. Sometimes a CTI report mentions an attack pattern that closely resembles the title. However, we need the description at other times to identify the matching technique. For example, for MITRE ATT\&CK ID: Location Tracking, we have descriptions of the form \textit{``eSurv can track the device’s location"} for one malware and \textit{``Pallas tracks the latitude and longitude coordinates of the infected device"} for another malware.\footnote{\url{https://attack.mitre.org/techniques/T1430/}} By computing the similarity between the attack pattern title and description, we can correctly match against both.
Specifically, we compute a metric, the weighted distance  between the extracted phrase and an ATT\&CK ID $i$, as follows:
\begin{equation*}
    d_i = w_t cos (v_{phrase}, v_{title}^i) + (1-w_t) cos (v_{phrase}, v_{desc}^i)
\end{equation*}
Where $v_{phrase}$ is the vector embedding for the extracted attack phrase, $v_{title}^i$ and $v_{desc}^i$ are the vector embeddings for the $i^{th}$ ATT\&CK ID, respectively. $cos$ represents the cosine distance between two vectors $u, v$ computed as 

\begin{equation*}
    cos(u,v) = 1-\frac{u . v}{||u||_2 ||v||_2}
\end{equation*}
We iterate over all the different attack patterns present for a platform (66 techniques for mobile platforms and 196 for enterprise platforms) and find the ID with the smallest distance. We output this as the mapped ID if the distance is less than a threshold $\tau$. We identify the optimum value for $tau$ experimentally.    
\end{enumerate}

Unstructured threat reports can lead to variations in the description of the same attack pattern across different reports. This redundancy of information can impede the effectiveness of pattern prediction. To mitigate this issue, we map the extracted attack patterns to standardized ATT\&CK techniques, allowing us to have a fixed number of attack patterns in the knowledge graph. 
\par
Although our model is trained on mobile platform CTI reports, it can be utilized to extract attack patterns for other platforms. Our platform-agnostic algorithm can extract attack patterns as they appear in the text. The mapping steps can be changed to incorporate the appropriate list of attack patterns in MITRE for the target platform. In Section \ref{sec:usecase}, we demonstrate the effectiveness of our approach in extracting attack patterns from enterprise CTI reports.

\subsection{Adding Relationship to Concepts}
We train a relation classification model to determine the relationship between each pair of entities mentioned in the report. We only consider a pair of entities for relation extraction if a valid relationship may exist between them according to the adopted ontology~\cite{rastogi2020malont,christian2021ontology}. For example, we may have a relationship between a pair of entities of type \textsf{Malware} and \textsf{Application}, e.g., $\langle Malware, targets, Application\rangle$. However, we do not have a valid relationship type when two entities are of type \textsf{Application} and \textsf{Time}. Similar to NER, we use transformer-based models for the relation extraction task. Our approach incorporates entity information for relation classification \cite{wu2019enriching}. Given a text $s$ with a pair of entities $e_1$ and $e_2$, we introduce four tokens that capture the position information of the entities. Consider the example:
\newline
\small
\textit{``Cerberus is capable of generating an instance of TeamViewer on mobile.''}
\normalsize
\newline
\textit{where}, $e_1$ is \textit{Cerberus} and $e_2$ is \textit{TeamViewer}. The formatted sentence will be: 
\textit{[CLS] $\langle$e1$\rangle$ Cerberus $\langle$/e1$\rangle$ is capable of generating an instance of $\langle$e2$\rangle$ TeamViewer $\langle$/e2$\rangle$ on mobile.} 

\textit{[CLS]} is the unique start of sequence token for the BERT model. We concatenate the hidden layer representation for the start position of both entities and generate the final vector embedding. We pass the vector through a couple of fully connected layers to predict the relation type between the entities. Once the triples are generated from the large corpus, the threat intelligence knowledge graph is ready for querying.

\subsection{Querying \textsf{LADDER}} \label{queryLadder}

\textbf{Knowledge Graph. } The threat intelligence graph is a directed knowledge graph, \textsf{KG} = \{{$\mathcal{E},\mathcal{R},\mathcal{T}$}\}, where {$\mathcal{E},\mathcal{R}$} and {$\mathcal{T}$} indicate the sets of entities, relations, and triples, respectively \cite{ji2021survey}. Each triple $\langle e_{head},r,e_{tail}\rangle \in\mathcal{T}$ indicates that there is a relationship \textit{r} $\in \mathcal{R}$ between $e_{head} \in \mathcal{E}$ and $e_{tail}\in \mathcal{E}$. \textsf{KG} link prediction is the task of predicting the best candidate for a missing entity. Formally, the task of entity-prediction is to predict the value for \textbf{ $e_{head}$ } given $\langle ?,r,e_{tail}\rangle$ or \textbf{$e_{tail}$} given $\langle e_{head},r,?\rangle$, where ``\textbf{?}" indicates a missing entity (head or tail).
\par
\textbf{Vector Embeddings}: All triples are mathematically represented by three vectors,  $\mathbf{e_1, e_2} \in \mathbb{R}^{d_e}$, $r \in \mathbb{R}^{d_r}$, where $d_e$ and $d_r$ are the embedding dimensions of entities and relations, respectively and have relatively low (e.g., 30-200) dimensional vector space embeddings~\cite{wang2017knowledge,balavzevic2019tucker}. Embeddings preserve information about the structure and key features of the triple. The embeddings for all the triples in the \textsf{KG} involve factorization of co-occurrence-based tensors \cite{balavzevic2019tucker}, which leads to reducing the dimensionality of the triples when creating embeddings for entities and relations in the \textsf{KG}. We use the $f_\phi$ notation for the scoring function of each triple $\langle e_{head},r,e_{tail}\rangle \in\mathcal{T}$. The scoring function measures the plausibility of a fact in $\mathcal{T}$, based on translational distance or semantic similarity \cite{li2020survey}.
% For the entities and relationship $e_{head}$, $e_{tail}, r$, we use the boldface \textbf{h, t, r} to indicate their low-dimensional embedding vectors, respectively. The symbols $n_e$, $n_r$, and $n_t$ denote the number of entities, relations, and triples of the KG. 
\par
\textbf{Querying \textsf{KG}. }With the threat intelligence graph, \textsf{KG} built, security analysts can query the graph where the query is posed in a triple format. Querying a knowledge graph is related to knowledge graph link prediction. Using the link prediction approach, \textsf{LADDER} predicts $e_{tail}$ by learning a scoring function. Entity prediction for \textsf{KG} follows TuckER \cite{balavzevic2019tucker} since it outperforms traditional link prediction models \cite{rastogi2022tinker}. TuckER is a linear model based on tucker decomposition \cite{tucker1966some} of entity embedding matrix and relational embedding matrix in a knowledge graph. We also evaluate the accuracy and recommend ranking inferred entities \cite{balavzevic2019tucker}.

\section{Experiments and Results}
\subsection{Information Extraction}

\begin{table}[]
\centering
\caption{Results for NER using different transformers (bold indicates the best result)}
\label{tab:ner-all}
\resizebox{0.7\columnwidth}{!}{%
\begin{tabular}{@{}llcc@{}}
\toprule
\textbf{Model} & \textbf{Precision} & \multicolumn{1}{l}{\textbf{Recall}} & \multicolumn{1}{l}{\textbf{F1-score}} \\ \midrule
BERT-base & 73.34 & 77.88 & 75.14 \\
BERT-large & 75.30 & 79.23 & 77.12 \\
RoBERTa-base & 41.55 & 41.01 & 40.84 \\
RoBERTa-large & 35.95 & 36.23 & 35.49 \\
XLM-RoBERTa-base & 75.32 & 79.06 & 76.98 \\
XLM-RoBERTa-large & \textbf{76.97} & \textbf{81.57} & \textbf{78.98} \\ \bottomrule
\end{tabular}%
}
\end{table}

% \usepackage{booktabs}
\begin{table}[]
\tiny
\centering
\caption{Entity extraction result for different classes using XLM-RoBERTa-large model}
\label{tab:ner-class}
\resizebox{0.7\columnwidth}{!}{%
\begin{tabular}{@{}lccc@{}}
\toprule
\textbf{Class} & \multicolumn{1}{l}{\textbf{Precision}} & \multicolumn{1}{l}{\textbf{Recall}} & \multicolumn{1}{l}{\textbf{F1-score}} \\ \midrule
Malware & 78.45 & 83.08 & 80.70 \\
MalwareType & 65.64 & 87.18 & 74.89 \\
Application & 70.13 & 73.26 & 71.66 \\
OS & 89.95 & 96.24 & 92.99 \\
Organization & 73.68 & 74.12 & 73.90 \\
Person & 88.24 & 75.00 & 81.08 \\
ThreatActor & 58.33 & 37.84 & 45.90 \\
Time & 85.51 & 89.39 & 87.41 \\
Location & 93.55 & 89.92 & 91.70 \\
Average & 76.97 & 81.57 & 78.98 \\ \bottomrule
\end{tabular}
}
\end{table}

We used the PyTorch deep learning framework to implement our models for information extraction and subsequent tasks. To achieve high performance, we fine-tuned the pre-trained transformer models from Huggingface's transformers library. For the Named Entity Recognition (NER) models, we set the sequence length to 128 and trained the models for 20 epochs, with 32 samples per mini-batch. We used the AdamW optimizer with a learning rate of 1e-5 for the base (BERT-base, RoBERTa-base, XLM-RoBERTa-base) and 1e-6 for large models (BERT-large, RoBERTa-large, XLM-RoBERTa-large). To avoid the repetition of reports written on the same malware, we split the annotated datasets based on the malware under discussion. Out of 36 malware, the training dataset comprised 26, and the validation and test datasets were five malware each. In total, there were 104 large CTI reports in the training split, 21 in the validation split, and 25 in the test split.

\subsubsection{Entity Extraction. }
The results for the named entity extraction task are shown in Table \ref{tab:ner-all}. XLM-RoBERTa large model achieves the best performance for this task with an average F1 score of 78.98\%. Class-specific results for this model are shown in Table \ref{tab:ner-class}. Classes with more samples in the training data generally produce better results. \textsf{Operating System} and \textsf{Location} classes yield better results than the other classes as they exhibit less variation in form. Performance reduces for \textsf{Application} and \textsf{Organization} as they have some overlap between them (e.g., Facebook and Twitter can be both \textsf{Application} and  \textsf{Organization}). The most challenging class is the \textsf{ThreatActor} because it is usually an \textsf{Organization} or \textsf{Person} with malicious intent. So, this requires a thorough understanding of the context to detect them correctly. Having a limited input length means this may not be possible to infer from a single sentence. There are also not enough samples for this class in the training data, further reducing the performance.

\begin{table}[]
\tiny
\centering
\caption{Result for the relevant sentence extraction subtask for attack pattern extraction}
\label{tab:sent-cls}
\resizebox{0.7\columnwidth}{!}{%
\begin{tabular}{@{}lccc@{}}
\toprule
\textbf{Model} & \textbf{Precision} & \textbf{Recall} & \textbf{F1-score} \\ \midrule
BERT-base & 86.50 & 85.20 & 85.84 \\
BERT-large & 86.06 & 85.80 & 85.93 \\
RoBERTa-base & 87.42 & 86.10 & 86.76 \\
RoBERT-large & \textbf{89.22} & \textbf{90.03} & \textbf{89.62} \\
XLM-RoBERTa-base & 83.00 & 88.52 & 85.67 \\
XLM-RoBERTa-large & 84.73 & 88.82 & 86.73 \\ \bottomrule
\end{tabular}%
}
\end{table}

% \usepackage{booktabs}
% \usepackage{graphicx}
\begin{table}[]
\tiny
\centering
\caption{Result for the attack phrase extraction subtask for attack pattern extraction}
\label{tab:ent-ext}
\resizebox{.7\columnwidth}{!}{%
\begin{tabular}{@{}lccc@{}}
\toprule
\textbf{Model} & \textbf{Precision} & \textbf{Recall} & \textbf{F1-score} \\ \midrule
BERT-base & 87.67 & 90.55 & 89.09 \\
BERT-large & 87.74 & 87.81 & 87.78 \\
RoBERTa-base & 88.53 & 90.12 & 89.32 \\
RoBERT-large & \textbf{89.19} & \textbf{92.14} & \textbf{90.64} \\
XLM-RoBERTa-base & 86.82 & 90.72 & 88.73 \\
XLM-RoBERTa-large & 88.55 & 91.77 & 90.13 \\ \bottomrule
\end{tabular}%
}
\end{table}

\subsubsection{Attack Pattern Extraction}
We use the same malware split in the NER task for attack pattern extraction. In a CTI report, usually, there are more sentences without an attack pattern than those that do. So, we randomly sample an equal number of negative sentences to balance the dataset for the sentence classification task. We fine-tune the transformer models for sentence classification and attack phrase extraction subtasks. We use a sequence length of 256 to train these models. To train the models, we use a mini-batch size of 32, and the optimal learning rate is chosen from [1e-5, 5e-5, 1e-6] using the validation set. We train the sentence classification and attack phrase extraction models for 20 and 30 epochs, respectively. We use Adam~\cite{kingma2014adam}as the optimizer for learning the model parameters.
\par
Table \ref{tab:sent-cls} shows the binary sentence classification task results. We achieve greater than 85\% F1-score for all the models, with RoBERTa large performing best with an average F1-score of 89.62\%. Table \ref{tab:ent-ext} shows the result for the attack phrase extraction task. Again, the RoBERTa-large model achieves the best result with an average F1-score of 90.64\%. These results indicate that our algorithm can effectively identify the relevant sentences for attack patterns and accurately extract relevant parts. To learn the optimal parameter values $w_t$ and $\tau$ required for mapping to MITRE ID, we manually map 80 randomly selected attack pattern descriptions annotated in our training corpus with their corresponding ATT\&CK ID. The optimum values for the two parameters are $0.4$ and $0.6$, respectively.

% \usepackage{graphicx}
\begin{table}[t]
% \tiny
\centering
\caption{Result for relationship extraction}
\label{tab:rel-ext}
\resizebox{.7\columnwidth}{!}{%
\begin{tabular}{lrrr}
\hline
\textbf{Model} & \multicolumn{1}{c}{\textbf{Precision}} & \multicolumn{1}{c}{\textbf{Recall}} & \multicolumn{1}{c}{\textbf{F1-score}} \\ \hline
BERT-base & 93.75 & 92.22 & 92.60 \\
BERT-large & \textbf{93.78} & \textbf{92.46} & \textbf{92.62} \\
RoBERTa-base & 81.66 & 83.88 & 82.27 \\
RoBERT-large & 77.47 & 81.06 & 77.97 \\
XLM-RoBERTa-base & 81.77 & 83.86 & 81.74 \\
XLM-RoBERTa-large & 72.64 & 78.69 & 75.29 \\ \hline
\end{tabular}%
}
\end{table}

% \usepackage{booktabs}
% \usepackage{graphicx}
\begin{table}[t]
\tiny
\centering
\caption{Class-specific results for relationship extraction}
\label{tab:rel-cls}
\resizebox{.7\columnwidth}{!}{%
\begin{tabular}{@{}lccc@{}}
\toprule
\textbf{Class} & \textbf{Precision} & \textbf{Recall} & \textbf{F1-score} \\ \midrule
noRelation & 100.0 & 100.0 & 100.0 \\
isA & 98.6 & 97.3 & 98.0 \\
targets & 96.3 & 88.1 & 92.0 \\
uses & 46.2 & 33.3 & 38.7 \\
hasAuthor & 87.5 & 93.3 & 90.3 \\
has & 100.0 & 70.0 & 82.4 \\
variantOf & 100.0 & 46.2 & 63.2 \\
hasAlias & 26.3 & 71.4 & 38.5 \\
indicates & 75.9 & 97.6 & 85.4 \\
discoveredIn & 100.0 & 100.0 & 100.0 \\
exploits & 100.0 & 50.0 & 66.7 \\ \bottomrule
\end{tabular}%
}
\end{table}

\subsection{Relation Extraction}

When extracting relationships from threat reports, evaluating each pair of entities and inferring the relationship between them is essential. However, many pairs of entities may not have a meaningful relation in the context, even if they are valid according to the ontology. To account for such cases, we add the \textit{NoRelation} class that predicts the absence of a relation between the entities under consideration. We randomly sample such plausible entities from the annotated CTI reports and use an 80:20 split for training and inference. Since the CTI reports may have a relation between entities that are far apart in the reports, we increase the sequence length to 512. Due to GPU memory constraints, we use a mini-batch size of 8 for the large and 16 for the small transformer models, respectively. We determine the optimal learning rate from [1e-5, 5e-5, 1e-6] and train the models for ten epochs using the AdamW optimizer.
\par
Table \ref{tab:rel-ext} shows the model performance for the relation extraction task. The BERT-based model outperforms the other two models for this task. The BERT-large model shows the highest performance with a 92.62\% average F1-score. Table \ref{tab:rel-cls} shows the results for specific classes. Attack patterns are part of a single relation type (\textit{Malware uses AttackPattern}), therefore not included in the relation extraction. Consequently, we can infer the relation for the attack pattern once the malware under discussion is identified. While \textit{discoveredIn} performs well for the annotated corpus because the reports were cleaned and mostly did not contain redundant time information on malware discovery. Among other classes, \textit{uses} and \textit{hasAlias} have the worst performance. Context is challenging with $uses$ as it may get confused with \textit{targets} as both contain the same type of head-tail entity pairs (e.g., \textsf{Malware} and \textsf{Application}. Relationship between two malware (\textit{variantOf} and \textit{hasAlias}) is challenging to detect since they may be expressed at a distant position in the report. It is important to note that these results are obtained from the manually cleaned test dataset, and the performance declines with more noisy texts.

\subsection{Threat Intelligence Knowledge Graph}

% \usepackage{booktabs}
% \usepackage{graphicx}
% \begin{table}[]
% \tiny
% \centering
% \caption{Number of triples in different datasets for the link prediction task}
% \label{tab:tucker-data}
% \resizebox{.4\columnwidth}{!}{%
% \begin{tabular}{@{}lcc@{}}
% \toprule
% \textbf{Dataset} & \textbf{TestSet 1} & \textbf{TestSet 2} \\ \midrule
% Test & 756 & 1209 \\
% $KG_1$ & 2265 & 1812 \\
% KG2 & 34830 & 34377 \\
% KG3 & 52648 & 52195 \\ \bottomrule
% \end{tabular}%
% }
% \end{table}

% {
% \color{cyan}

% \begin{table}[]
% \tiny
% \centering
% \caption{Number of triples in different knowledge graph datasets for the link prediction task (VT: triples generated using VirusTotal)}
% \label{tab:tucker-data}
% \resizebox{.75\columnwidth}{!}{%
% \begin{tabular}{@{}llcc@{}}
% \toprule
% \textbf{Dataset} & \textbf{Data Source}         & \textbf{TestSet 1} & \textbf{TestSet 2} \\ \midrule
% Test             & Hand Annotated               & 756                & 1209               \\
% $KG_1$              & Hand Annotated               & 2265               & 1812               \\
% KG2              & Hand Annotated+LADDER        & 34830              & 34377              \\
% KG3              & Hand Annotated + LADDER + VT & 52648              & 52195              \\ \bottomrule
% \end{tabular}%
% }
% \end{table}

% }

The trained information and relation extraction models allow us to generate triples from new CTI reports. To extract the concepts, we combine prediction from the best-performing NER model as well as heuristics to extract the concepts. We perform some post-processing to remove noisy entities extracted by the approach. For \textsf{Malware} and \textsf{ThreatActors}, we exclude them if they are predicted as a different class elsewhere or only mentioned once. For example, organizations like \textit{ThreatFabric} are sometimes classified as \textit{ThreatActor} instead of \textit{Organization}. We do not use the entity node for relation extraction in such cases. Next, we apply the relation extraction model with the best performance measure to determine the relationship between entity pairs following the ontology. We extract and map attack patterns to their corresponding MITRE IDs using our proposed \textsf{TTPClassifier} algorithm. We identify the malware under discussion per the CTI report (the most frequently mentioned malware, if any) and associate the relationship with the malware and the attack patterns.

% Since there can be incorrect triples extracted due to noisy text and incorrect predictions from the model, we perform post-processing to reduce incorrect triples. To do this, we manually identify all the incorrect \textsf{Malware} and \textsf{ThreatActor} extracted, i.e., false positives, and remove the triples involving them.

% We also include triples from Virus Total (VT) to create a larger knowledge graph to evaluate link prediction performance in the presence of more IoCs. VT provides academic API to access intelligence on a database of several million malware samples. We use the VT API V3.0 \cite{vt-website} to extract IoCs (contacted domains and IP addresses) and permission requests for each malware hash collected from a larger corpus of OpenCTI reports. We map each permission to attack pattern and MITRE ID using \textsf{TTPClassifier} and evaluate attack pattern prediction from querying the larger knowledge graph.

\subsection{Inferring Entities}

We create two test sets from the hand-annotated documents with varying triples for the prediction task. Entities in these triples belong to classes \textsf{Malware}, e.g., \textsf{AttackPattern}, \textsf{Location}, \textsf{Application}, \textsf{Organization} for testing. TestSet-1 consists of 25\% of the annotated triples, and $TestSet_2$ consists of 40\% of the triples. We experiment with three knowledge graphs for the prediction task; $KG_1$ has the remaining triples in hand-annotated CTI reports, and $KG_2$ consists of the triples generated using \textsf{LADDER} from the 12,000 documents. We train TuckER to predict the tail entities, employing 50 embedding dimensions with a mini-batch size of 64. The model is trained for 1000 iterations with an initial learning rate of 0.001.  
\par
\textbf{Evaluation Criteria. }To rank the performance of the prediction task, particularly in the context of knowledge graph prediction, we use evaluation criteria, including Mean Rank, Mean Reciprocal Rank (MRR), and Hits@\textit{n}. These are calculated from the ranks of all true (actual) test triples that TuckER returns. MRR is the average inverse of the ranks of all the true test triples. Hits@\textit{n} denotes the percentage of test-set ranking where a true triple is ranked within the top \textit{n} positions of the ranking. Higher scores are considered better.

% \usepackage{booktabs}
% \usepackage{multirow}
% \usepackage{graphicx}
\begin{table}[]
\centering
\caption{Inference (link prediction) results for different training and test datasets}
\label{tab:link-all}
\resizebox{1\columnwidth}{!}{%
\begin{tabular}{@{}lcccccccc@{}}
\toprule
\multirow{2}{*}{KG} & \multicolumn{4}{c}{$TestSet_1$} & \multicolumn{4}{c}{$TestSet_2$} \\ \cmidrule(r){2-5} \cmidrule(lr){6-9}
 & Hits@3 & Hits@10 & Hits@30 & MRR & Hits@3 & Hits@10 & Hits@30 & MRR \\ \midrule
$KG_1$ & 0.209 & 0.365 & 0.497 & 0.186 & 0.090 & 0.195 & 0.322 & 0.093 \\
$KG_2$ & 0.221 & 0.353 & 0.516 & 0.211 & 0.215 & 0.359 & 0.501 & 0.203 \\
% KG3 & 0.202 & 0.353 & 0.507 & 0.190 & 0.204 & 0.356 & 0.498 & 0.193 \\
\bottomrule
\end{tabular}%
}
\end{table}

% \usepackage{booktabs}
% \usepackage{multirow}
% \usepackage{graphicx}
\begin{table}[]
% \tiny
\centering
\caption{Class-specific inference (link prediction) results.}
\label{tab:tucker-cls}
\resizebox{1\columnwidth}{!}{%
\begin{tabular}{@{}lccccccccc@{}}
\toprule
\multirow{2}{*}{\textbf{Class}} & \multirow{2}{*}{\textbf{KG}} & \multicolumn{4}{c}{\textbf{$TestSet_1$}} & \multicolumn{4}{c}{\textbf{$TestSet_2$}} \\ \cmidrule(l){3-6} \cmidrule(l){7-10}
 &  & Hits@3 & Hits@10 & Hits@30 & MRR & Hits@3 & Hits@10 & Hits@30 & MRR \\ \midrule
\multirow{2}{*}{AttackPattern} & $KG_1$ & 0.354 & 0.634 & 0.847 & 0.314 & 0.212 & 0.441 & 0.657 & 0.210 \\
 & $KG_2$ & 0.444 & 0.700 & 0.940 & 0.420 & 0.453 & 0.694 & 0.936 & 0.415 \\
\multirow{2}{*}{Location} & $KG_1$ & 0.042 & 0.096 & 0.205 & 0.048 & 0.018 & 0.033 & 0.096 & 0.024 \\
 & $KG_2$ & 0.036 & 0.096 & 0.247 & 0.044 & 0.018 & 0.092 & 0.225 & 0.034 \\
\multirow{2}{*}{Application} & $KG_1$ & 0.100 & 0.165 & 0.230 & 0.086 & 0.032 & 0.094 & 0.178 & 0.040 \\
 & $KG_2$ & 0.026 & 0.083 & 0.126 & 0.030 & 0.040 & 0.102 & 0.129 & 0.034 \\ \bottomrule 
\end{tabular}%
}
\end{table}

\textbf{Results. }Table \ref{tab:link-all} presents the results for the inference task. $TestSet_1$ performs similarly using the different training datasets for all Hits@(n) values. However, for $TestSet_2$, $KG_1$ performs worse than $KG_2$, with a difference of 16.4\% Hits@10 between $KG_1$ and $KG_2$. This suggests that the additional triples obtained from a larger knowledge graph enable the model to make better predictions. $KG_2$ performs similarly on both the test datasets suggesting better generalizability. We show class-specific prediction results in Table \ref{tab:tucker-cls} using KG1 and KG2 for three different tail entities - \textsf{AttackPattern}, \textsf{Location} and \textsf{Application} where the head entity is \textsf{Malware}. The most promising result is obtained for predicting \textsf{AttackPattern}, primarily from having 66 unique attack patterns. At the same time, the number of possible tail entities is much larger for relations involving other classes. 
This result implies that malware with similar properties may exhibit similar attack patterns. Similar to the aggregate result above, we observe no significant drop in performance for the two test datasets for $KG_2$. $KG_1$ sees a significant drop.

\subsection{Comparison with state-of-the-art for TTP Classifier}

We compare our attack pattern extraction with TTPDrill \cite{husari2017ttpdrill} and AttackKG \cite{li2022attackg} as they are the closest to \textsf{LADDER}, although there are differences that we discuss in the related work section. We use the open-source implementation of TTPDrill and AttackKG in our evaluation. Since TTPDrill and AttackKG provide models and patterns for the enterprise platform, we use the same for evaluation. We update the third step of our proposed \textsf{TTPClassifier} and match extracted phrase against the attack patterns listed on MITRE ATT\&CK for enterprise \footnote{\url{https://attack.mitre.org/techniques/enterprise/}}. Even though we trained our sentence classification and phrase extraction models for attack patterns on CTI gathered on mobile platforms, our evaluations also show high accuracy when testing CTI for other platforms since the semantic style for describing attack patterns is the same. This is because the description of the techniques follows a similar pattern in written texts.

% \usepackage{booktabs}
% \usepackage{graphicx}
\begin{table}[t]
\centering
\caption{Comparison of attack pattern extraction with other methods (TP: true positives, FN: False Negatives, FP: False positives)}
\label{tab-atk-pattern-result-comparison}
\resizebox{0.7\columnwidth}{!}{%
\begin{tabular}{@{}lcccccc@{}}
\toprule
\textbf{Method} & \textbf{TP} & \textbf{FN} & \textbf{FP} & \textbf{Precision} & \textbf{Recall} & \textbf{F1-score} \\ \midrule
\textbf{MITRE} & 38 & 27 & 0 & \textbf{1.00} & 0.58 & \textbf{0.74} \\
\textbf{TTPDrill\cite{husari2017ttpdrill}} & 22 & 43 & 231 & 0.09 & 0.34 & 0.14 \\
\textbf{AttackKG\cite{li2022attackg}} & 12 & 53 & 85 & 0.12 & 0.18 & 0.15 \\
\textbf{TTPClassifier} & 41 & 24 & 22 & 0.65 & \textbf{0.63} & 0.64 \\ \bottomrule
\end{tabular}%
}
\end{table}

When analyzing the attack patterns listed in the MITRE ATT\&CK website for malware, \textit{we often found that not all attack patterns reported in a CTI report are present. This indicates the difficulty of this task, even for human annotators}. In order to have a fair comparison, we create ground truth annotation from threat reports listed on the MITRE ATT\&CK website for five different malware containing 9360 tokens. We show the result in Table \ref{tab-atk-pattern-result-comparison}. The attack patterns listed on the MITRE website have the overall best F1 score. Even though we did not find any false positives, 27 out of the 65 attack patterns we identified were not listed. This suggests that it is very likely that security analysts may miss some attack patterns when reading long, detailed CTI reports. Our proposed TTPClassifier achieves better recall than those listed on the MITRE website. TTPDrill and AttackKG achieve similar F1 scores, with TTPDrill showing better recall but having a lot of false positives. Since both approaches use template matching based approach generated from attack pattern description, they are ill-equipped to filter out irrelevant parts of large documents, which results in many false positives. Another issue with the template-matching-based approach is that it is challenging to identify a novel attack pattern when there is not enough example pattern available for an attack pattern. However, our machine-learning-based approach can mitigate this issue and identify new or emerging attack patterns.

\subsection{Error Analysis} 
Since TTPClassifier is a multi-step algorithm, errors in an earlier stage may influence the result of the next stage. For example, if a sentence is misclassified as irrelevant, attack patterns present in that step will not be captured in the subsequent steps. However, a CTI report often describes the same attack pattern in multiple places. As a result, the attack pattern descriptions missed from one part of the report may be captured by another. This effect is further lessened when attack patterns are aggregated from multiple threat reports. Another case of an error that may be introduced in the first step is erroneously classifying an irrelevant sentence as positive. However, since the second step of the algorithm is independent of the first, some of them may be reduced if the model cannot identify any descriptive phrase of attack pattern in the sentence. False positives introduced in the second step may not match any existing attack pattern description and may be omitted in the final output. There may be errors introduced in the last step of the algorithm; however, since the framework can provide accompanying descriptions, analysts can verify them. We show statistics of the errors introduced after each step of the algorithm in Table \ref{tab:err-propagation} for the analyzed reports. The number of false positives is usually additive unless false positives introduced in one step are removed in a later stage. Also seen is that most of the false negatives are introduced in the first step of the algorithm, while most false positives are introduced in the last stage. Since the reports used in the experiment are all for different malware, the effect of false negatives will be reduced when results are aggregated from multiple reports for the same malware.

\begin{table}[t]
\centering
\caption{The number of errors introduced after each algorithm step.}
\label{tab:err-propagation}
\resizebox{0.7\columnwidth}{!}{%
\begin{tabular}{@{}lcc@{}}
\toprule
\textbf{Step}                         & \textbf{FN} & \textbf{FP} \\ \midrule
\textbf{1. Relevant Sentence Extraction} & 23          & 8           \\
\textbf{2. Attack Phrase Extraction}     & 3           & 11          \\
\textbf{3. Mapping to MITRE ID}          & 7           & 22          \\ \bottomrule
\end{tabular}%
}
\end{table}

\section{Case Studies}\label{sec:usecase}

\subsection{Attack Pattern Extraction and Trend Analysis}
\label{usecase-ttp}
% One of our research questions is how reliably we can extract attack patterns from unstructured CTI reports. The results in preceeding section indicates the proposed method is able to extract attack patterns from CTI reports on different platforms. 
% \textcolor{red}{[Discuss them in related work?]} There has been some prior research on AttackPattern extraction from CTI. The work in \cite{husari2017ttpdrill} used ...

% % Please add the following required packages to your document preamble:
% % \usepackage{booktabs}
% % \usepackage{graphicx}
% \begin{table}[t]
% \centering
% \label{tab-atk-pattern-result-comparison}
% \caption{Comparison of attack pattern extraction with other methods (TP: true positives, FN: False Negatives, FP: False positives)}
% \label{tab:atk-pattern-compare}
% \resizebox{0.7\columnwidth}{!}{%
% \begin{tabular}{@{}lcccccc@{}}
% \toprule
% \textbf{Method} & \textbf{TP} & \textbf{FN} & \textbf{FP} & \textbf{Precision} & \textbf{Recall} & \textbf{F1-score} \\ \midrule
% \textbf{MITRE} & 38 & 27 & 0 & \textbf{1.00} & 0.58 & \textbf{0.74} \\
% \textbf{TTPDrill\cite{husari2017ttpdrill}} & 22 & 43 & 231 & 0.09 & 0.34 & 0.14 \\
% \textbf{AttackKG\cite{li2022attackg}} & 12 & 53 & 85 & 0.12 & 0.18 & 0.15 \\
% \textbf{TTPClassifier} & 41 & 24 & 22 & 0.65 & \textbf{0.63} & 0.64 \\ \bottomrule
% \end{tabular}%
% }
% \end{table}

% \usepackage{booktabs}
\begin{table*}[]
\centering
\caption{Example Attack Patterns extracted from a threat report using TTPClassifier for LitePower malware}
\label{tab:litepower}
\resizebox{0.9\textwidth}{!}{

\begin{tabular}{@{}lllcc@{}}
\toprule
\textbf{MITRE ID} & \textbf{Name}                     & \textbf{Description in Report}                                  & \textbf{ATT\&CK}          & \textbf{TTPClassifier}                        \\ \midrule
T1059             & Command and Scripting Interpreter & use a PowerShell script to execute commands                     & \checkmark & \checkmark \\
T1041             & Exfiltration Over C2 Channel      & send collected data, including screenshots, over its C2 channel & \checkmark & \checkmark                     \\
T1012             & Query Registry                    & checks for the registry keys added for COM hijacking            & \checkmark & \checkmark                     \\
T1113             & Screen Capture                    & takes system screenshots and saves them to \% AppData \%        & \checkmark & \checkmark                     \\
T1053             & Scheduled Task/Job                & creation of a legitimate scheduled task                         & \checkmark & \checkmark                     \\
T1518             & Software Discovery                & can identify installed AV software                              & \checkmark & x                        \\
T1082             & System Information Discovery      & list local drives and enumerate the OS architecture             & \checkmark & x                         \\
T1564             & Hide Artifacts                    & hide the main dropper spreadsheet                               & x                         & \checkmark                     \\
T1112             & Modify Registry                   & current user registry hive (HKCU)                               & x                         & \checkmark                     \\
T1588             & Obtain Capabilities               & download and deploy further malware                             & x                         & \checkmark                     \\ \bottomrule
\end{tabular}
}
\end{table*}

The results in the preceding section indicate that the proposed algorithm can effectively extract attack patterns from CTI reports on different platforms. In this case study, we examine attack patterns extracted for a Windows malware, LitePower~\cite{mitreAttack} in Table \ref{tab:litepower}. We extract 18 attack patterns from this CTI report \cite{litepower-securelist}. We list the five attack patterns shared between TTPClassifier and attack patterns listed for that malware on the MITRE website. TTPDrill failed to identify two example attack patterns: \textit{T1518 Software Discovery}, and \textit{T1082 System Information Discovery}. Interestingly, our algorithm extracted the relevant phrases for those patterns. The extracted phrase for T1518 was \textit{conducts system reconnaissance to assess the AV software installed and the user privilege}, which got mapped to T1497-- Virtualization/Sandbox Evasion. Upon further inspection, we noticed the phrase matched the sub-technique \textit{System Checks}, which starts with the description \textit{Adversaries may employ various system checks to detect.} This description was similar to the first half of the phrase \textit{conducts system reconnaissance}, which resulted in the match. The extracted phrase for the second attack pattern was \textit{volumeserialnumber List local disk drives}. The word \textit{volumeserialnumber} was part of a Table in the CTI. This phrase did not match with any attack pattern with high enough similarity. The closest match was T1619-- Cloud Storage Object Discovery. However, the description of this attack pattern contains a reference to another attack pattern \textit{File and Directory Discovery}, which explains its relatively higher similarity.

\begin{figure}[th]
\tiny
\centering
\resizebox{1\linewidth}{!}{%
\includegraphics[width=.5\textwidth]{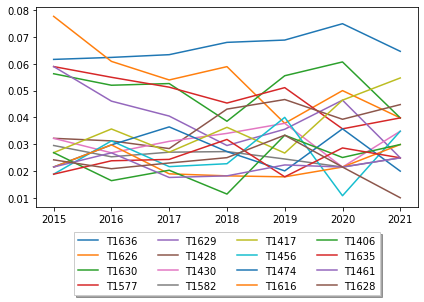}
\includegraphics[width=.5\textwidth]{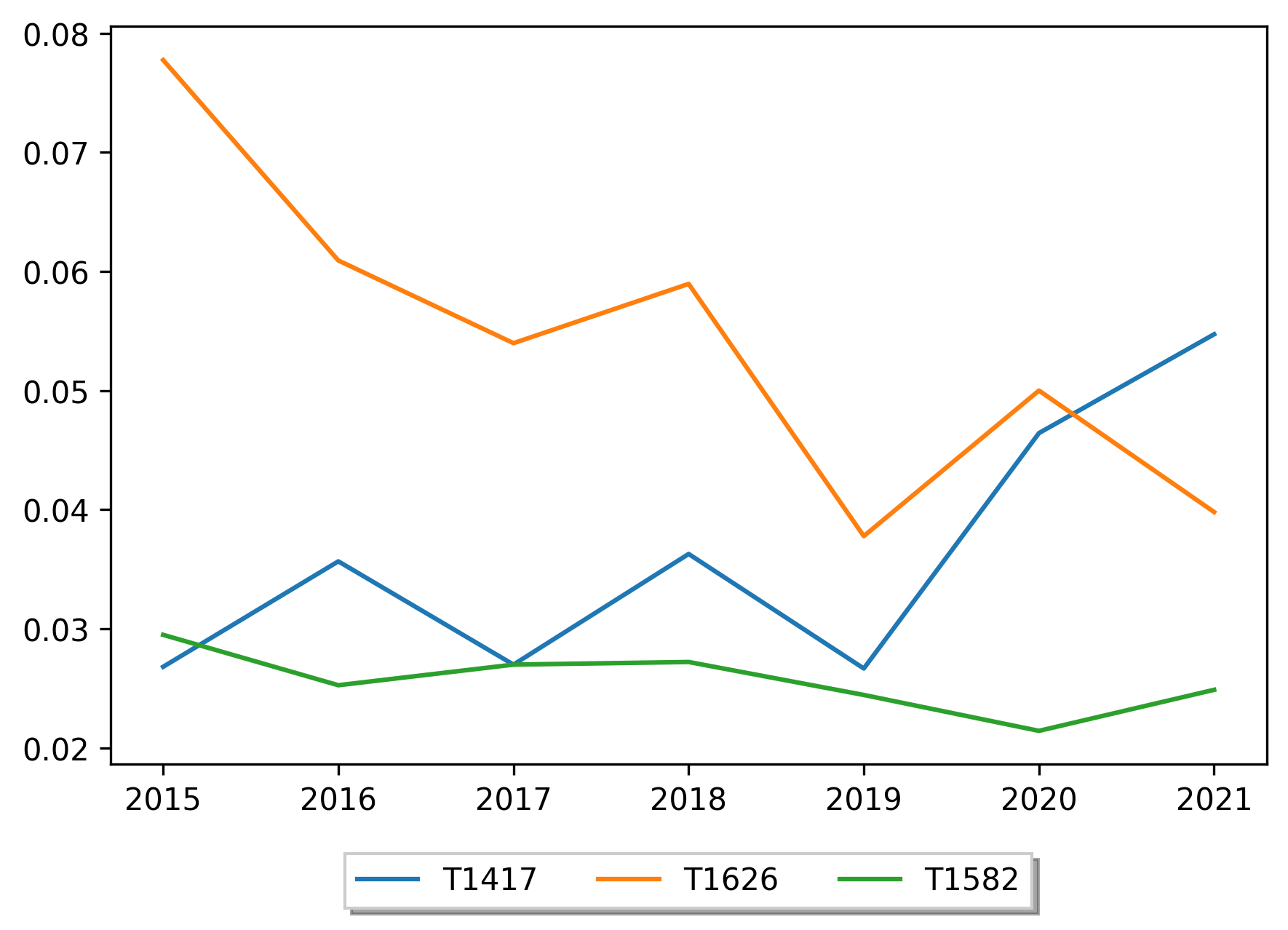}
}
\caption{Distributions of all (l) and three (r) attack techniques vs. time. The X-axis represents the year when an attack technique was observed in the CTI, and Y-axis represents the normalized count of that attack pattern (ratio of the count of an attack technique observed in that year to the total number of attack techniques observed in that year). }
\label{fig:attack_time_3}
\end{figure}

A significant advantage of our proposed approach is that we can extract relevant phrases from CTI even when they are not mapped correctly or do not have a unique mapping with MITRE attack patterns. An example of the latter is the absence of an attack pattern in MITRE mobile platforms for $Masquerading$, which is included for enterprise platforms \cite{masq}. However, we have noticed several Android malware exhibiting this attack pattern. One such malware is Ginp. As described in a threat report published by ThreatFabric \cite{tf-masq}, this malware was \textit{masquerading as a ``Google Play Verificator'' app}. Our approach can give an analyst the summarized version of the CTI report, including mentions of attack steps used by malware. We show \textbf{three example attack patterns} extracted by our algorithm but not listed on MITRE ATT\&CK. For example, the last attack pattern listed, T1588, describes that the malware can download and deploy further malware. These results suggest that the proposed algorithm can alleviate human labor when analyzing cyber threat reports. 
\newline
\textit{Large Scale Malware Behavior Analysis: } The proposed \textsf{TTPClassifier} is used to extract attack patterns for 433 malware instances found in 12K threat reports to perform attack pattern trend analysis. We only count unique attack patterns for the same malware if they are mentioned in multiple reports and obtain 3159 attack patterns between 2015 and 2021. 
We map the attack patterns to MITRE attack technique IDs and plot the distribution of attack IDs against time, see Figure \ref{fig:attack_time_3}(r) for three different trends. Refer to Figure \ref{fig:attack_time_3}(l) for a plot on all other attack techniques. The trend analysis shown in Figure \ref{fig:attack_time_3}(l) and Figure \ref{fig:attack_time_3}(r) are based on the CTI sources we have analyzed and are not representative of attack patterns deployed by malware in the wild.

% \begin{figure}[th]
% \tiny
% \centering
% \resizebox{0.8\linewidth}{!}{%
% \includegraphics[width=.45\textwidth]{figures/16.png}
% }
% \caption{Distributions of attack technique vs. time. The X-axis represents the year when an attack technique was observed in the CTI, and Y-axis represents the normalized count of that attack pattern (ratio of the count of an attack technique observed in that year to the total number of attack techniques observed in that year). }
% \label{fig:attack_time}
% \end{figure}

%\textcolor{blue}{no. of malware that used this attack pattern are NN.}
In Figure \ref{fig:attack_time_3}(r), we observe an upward trend of T1417 (Input Capture) with peak usage in 2021. This attack technique encompasses any methods an adversary uses to steal the application credentials of users. Adversaries can use keylogging or GUI capture methods to steal user input. For instance, Anubis malware has a keylogger that works in every application installed on the device \cite{anubis}. Over the years, a potential reason for the increase in this attack technique is the increased use of mobile-based digital solutions like banking, finance, and shopping. We expect this attack technique to become more prevalent in the coming years. We observe a downward trend of T1626 (abuse elevation control mechanism). T1626 encompasses methods an adversary uses to grant themselves high-level permissions in a device. For example, Red Alert 2.0 malware can request device administrator permissions \cite{redalert}. One reason for the reduction in this attack technique could be that users have become more cautious of applications that ask for device administration requests. Also, Android OS 7+ has introduced changes that make abuse of administrator privilege more difficult. As more devices update to the latest OS, adversaries will find it more difficult to manipulate users to gain elevated access to devices and use them for malicious purposes. T1582 (SMS Control) exhibits an almost flat trend over the years, with a slight dip and rise in multiple years. This attack ID represents the SMS control technique where an adversary deletes, alters, or sends SMS messages without user permission. For example, Anubis malware can send, receive, and delete SMS messages from a user's device \cite{anubis}. This trend suggests that SMS phishing remains as widespread on the mobile platform as before.

\subsection{Threat Hunting}\label{sec:threathunt}
\begin{figure}[t]
\centering
\includegraphics[width=.45\textwidth]{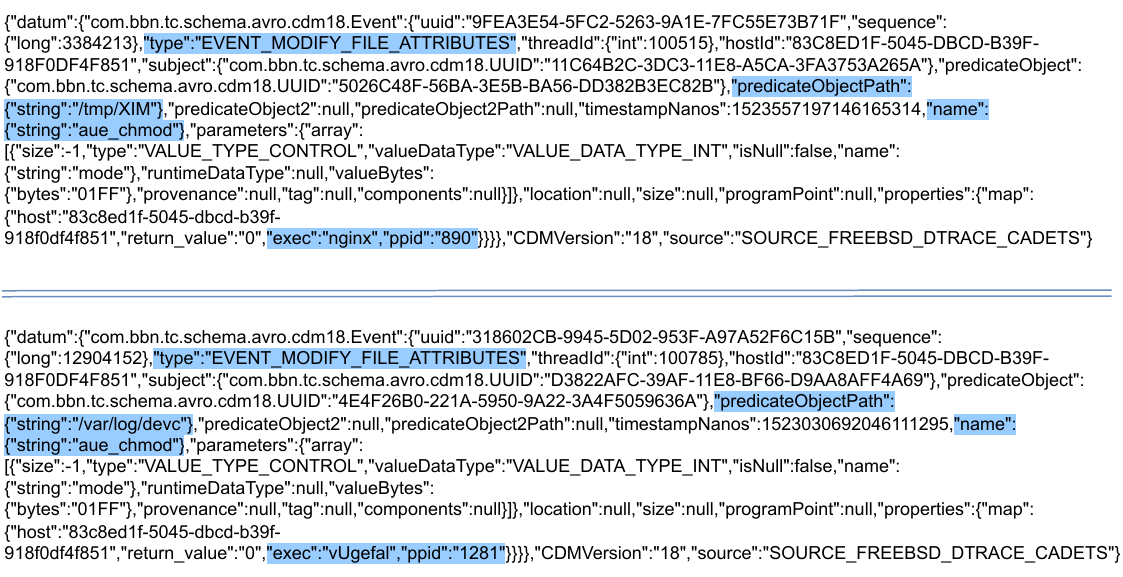}
\caption{2 actual logs for attack pattern T1222--	File and Directory Permissions Modification. Blue highlights relevant part(zoom for best view).}
\label{fig:flog-image}
\end{figure}

Automating the process of attack pattern extraction can assist in threat hunting and protection against APT campaigns. Correlating attack patterns extracted from CTI with kernel logs can pinpoint an attacker's activities. However, the same APT campaign may manifest differently in different settings due to differences in OS, targeted applications, and threat variations. As a result, relying on IoCs for precise threat hunting is unreliable since attackers can modify them to evade detection \cite{landauer2019framework}. We illustrate this with sample logs from the DARPA Transparent Computing Engagement 3 dataset in Figure \ref{fig:flog-image}. The logs collected in the figure display an attack on a  FreeBSD server that exploits an Nginx backdoor vulnerability. These logs exemplify the attack pattern  T1222-- File and Directory Permissions Modification which was used as a precursor to creating a new elevated process (attack pattern T1548). The accompanying ground truth CTI description refers to the \textit{ability to create a new elevated process}, with the interactions as $F1>elevate /tmp/XIM$. As we can see, there are variations in the logs due to the process ID, process name, and the file that is being modified. Consequently, instead of relying solely on matching the exact IoC, which may differ from what is described in a CTI report for a particular setting, it is more reliable to use high-level abstract information like attack patterns in conjunction with IoCs to identify attacks.

\textbf{Challenges. }Extracting attack patterns from kernel logs is beyond the scope of this study. There are a few ongoing open-source efforts for developing rules for MITRE attack pattern extraction from logs-- for example, the Sigma rules.\footnote{https://github.com/SigmaHQ/sigma} However, these rules do not yet adequately cover different attack patterns. Some rules are described in terms of specific tools or software, and when different software with the same functionality is used, they cannot be captured with the existing rules. Rules also differ based on the operating systems. For example, Out of 193 enterprise attack patterns in MITRE, there are sigma rules for 60 for Linux, resulting in a coverage of 31\%. As a result, many attack patterns identified through CTI may remain undetected in the log. Regardless, improvement in attack pattern extraction from logs can provide analysts with an efficient way of identifying specific APT attacks within kernel logs. This is one of our key objectives for future research.

\subsection{Attack Pattern Prediction}\label{tab:link-all}

Information extracted from a single CTI often lacks comprehensive information about a specific malware. Consolidating data from multiple reports allows us to bridge that gap and gain a more holistic perspective of any cyber attack. However, it is still possible that some specific malware characteristics may not be captured in the analyzed reports, resulting in missing information or gaps in the knowledge graph. We can find such missing information using knowledge graph link prediction, where the goal is to predict a missing tail entity given the head entity and the relation type. Another way to look into this is that as malware evolves, it can attempt new intrusion approaches resulting in attack patterns that have yet to be captured. We can use the link prediction task as a proxy to predict such attack patterns, which the malware may use in the near future.

We conduct a study of \textit{AttackPattern} prediction for two malware: Anubis and Flubot. Anubis is an Android banking trojan that was active from 2017 until late 2021 and employed a wide range of attack techniques \cite{anubisapk}. Flubot is another banking malware targeting Android users since the first quarter of 2021 and has been active ever since \cite{flubotapk}. We identified 33 unique attack patterns for Anubis and 23 for Flubot from the reports collected during these periods. We used triples from the $KG_2$ (see Table \ref{tab:link-all}) graph for the prediction task. We removed \textit{AttackPattern} triples for making a prediction (i.e. when predicting for Anubis, we remove triples of the form \textit{Anubis uses T1636}). Next, we used TuckER to predict the tail entities. For instance, we queried for the \textit{uses} relation, e.g., $\langle Anubis, uses, ?\rangle$. Table \ref{tab:usecase} shows the top 20 attack patterns predicted for both malware and confidence scores. As we include more predictions, it invariably leads to less confident outputs and results in more false positives. However, we see promising results for the top predictions. We get nine correct predictions for Anubis and seven correct predictions for Flubot out of the top 10. When considering the top 15, the number of correct predictions is 12 for both malware types. These findings suggest a strong correlation between different malware behavior. We can leverage a knowledge graph of this correlation for predicting unknown behavior, viz., inferring future attack patterns from available information. We can also use this to fill up missing information in existing knowledge graphs.

% \usepackage{booktabs}
% \usepackage{graphicx}
% \usepackage[table,xcdraw]{xcolor}}
\begin{table}[t]
\tiny
\centering
\caption{AttackPattern prediction for Anubis \& Flubot sorted by confidence (Green: Observed,Red: Not Observed Patterns.)}
\label{tab:usecase}
%\resizebox{\columnwidth}{!}{%
\resizebox{0.34\textwidth}{!}{%
\begin{tabular}{@{}cccc@{}}
\toprule
\multicolumn{2}{c}{Anubis} &  \multicolumn{2}{c}{Flubot} \\
\cmidrule(r){1-2} \cmidrule(lr){3-4}
 \textbf{Prediction} & \textbf{Confidence} & \textbf{Prediction} & \textbf{Confidence} \\
\midrule 
\colorbox{green!20}{T1636} & {0.522} & \colorbox{green!20}{T1636} & {0.743} \\
\colorbox{green!20}{T1626} & {0.410} & \colorbox{red!20}{T1630} & {0.661} \\

\colorbox{green!20}{T1630} & {0.406} & \colorbox{green!20}{T1577} & {0.652} \\

\colorbox{green!20}{T1577} & {0.385} & \colorbox{green!20}{T1626} & {0.646}\\

\colorbox{green!20}{T1629} & {0.354} & 
\colorbox{red!20}{T1428} & {0.592} \\

\colorbox{green!20}{T1428} & {0.351} & \colorbox{green!20}{T1629} & {0.580} \\

\colorbox{green!20}{T1430} & {0.274} & 
\colorbox{red!20}{T1430} & {0.508} \\

\colorbox{green!20}{T1417} & {0.258} & \colorbox{green!20}{T1417} & {0.455} \\

\colorbox{green!20}{T1582} & {0.251} & \colorbox{green!20}{T1474} & {0.444} \\

\colorbox{red!20}{T1406} & {0.244} & \colorbox{green!20}{T1628} & {0.423} \\

\colorbox{green!20}{T1456} & {0.229} & \colorbox{green!20}{T1406} & {0.419} \\

\colorbox{green!20}{T1474} & {0.226} & \colorbox{green!20}{T1582} & {0.406} \\

\colorbox{green!20}{T1616} & {0.226} & \colorbox{green!20}{T1635} & {0.386} \\

\colorbox{red!20}{T1628} & {0.224} & 
\colorbox{green!20}{T1456} & {0.386} \\

\colorbox{red!20}{T1461} & {0.208} & 
\colorbox{green!20}{T1616} & {0.380} \\

\colorbox{green!20}{T1635} & {0.205} & \colorbox{red!20}{T1625} & {0.371} \\

\colorbox{red!20}{T1625} & {0.195} & \colorbox{red!20}{T1461} & {0.359} \\

\colorbox{red!20}{T1404} & {0.169} & 
\colorbox{green!20}{T1634} & {0.335} \\

\colorbox{green!20}{T1639} & {0.163} & \colorbox{red!20}{T1639} & {0.323} \\
\bottomrule
\end{tabular}%
}
\end{table}

\subsection{Identifying Similar Malware and APT Groups}

We can use the information aggregated in the knowledge graph to identify similarities between entities, such as malware and threat actors. When a new malware emerges, it is of interest to security analysts to recognize similar malware, as this can provide valuable insight into the malware behavior. Similarly, if a previous APT group launches a new campaign, it may not be apparent initially from the limited information. However, if we can identify past APT groups with similar characteristics, it can help implement preventive measures against such an attack. 

\begin{figure}[t]
\centering
\resizebox{0.8\linewidth}{!}{%
\includegraphics[width=.5\textwidth]{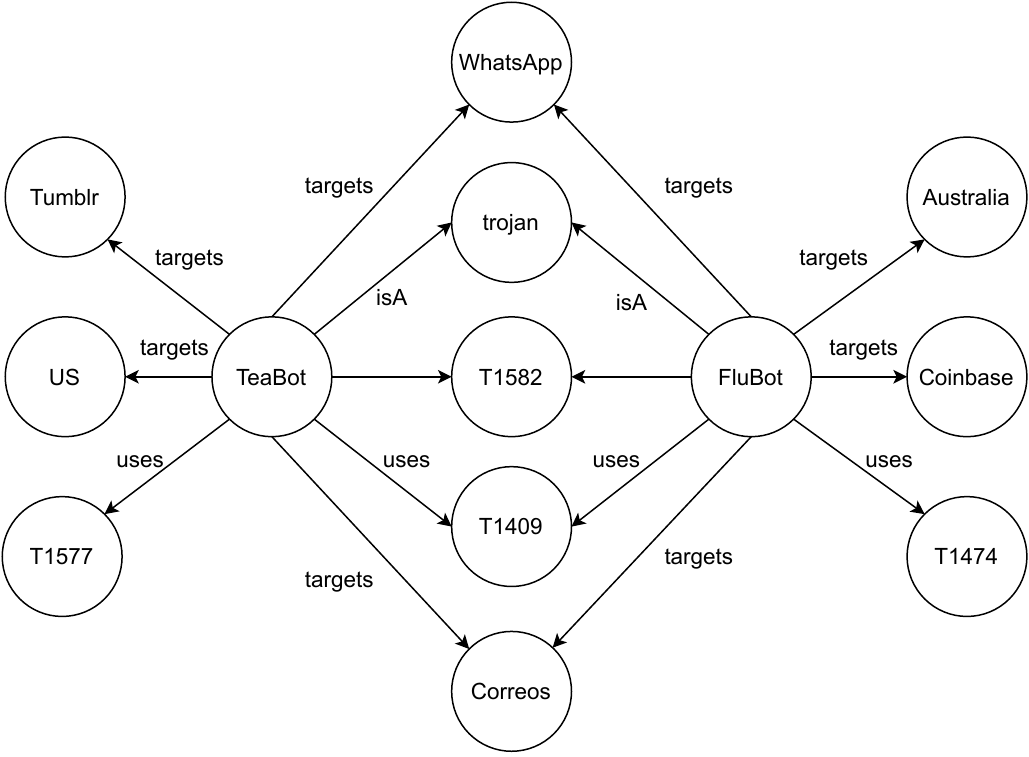}
}
\caption{Subgraph on similarity between FluBot \& TeaBot}
\label{fig:flu-teabot}
\end{figure}

In this study, we use the subgraph centered on an entity to find the similarity between malware and threat actors. We construct a set comprising all the adjacent nodes to malware consisting of entity and relation information. We use the Jaccard index \cite{jaccard} of these sets to measure the similarity between a pair of malware. We iterate over all the malware in our larger knowledge graph to identify the most similar one to a given malware. For example, using this approach, \textsf{TeaBot} was identified as the most similar malware to \textsf{FluBot}. Both are Android banking trojans that were active globally in 2021 \cite{flubot-teabot}. We show some of the shared nodes between the two malware in Figure \ref{fig:flu-teabot}. As we can see, both are Android trojans that target banking apps (Correos). Both have the attack pattern-- ``SMS control mechanism'' (ATT\&CK ID-T1582) by which they take control of the user's SMS app to interrupt incoming messages and send messages to spread the malware. They also collect stored data from the user's application (MITRE ID-T1409) and target some social media apps like WhatsApp. Some nodes are not shared between the two malware; for example, although FluBot targeted Australia, TeaBot did not. TeaBot targeted the Tumblr app, which FluBot did not. 
\par
When identifying similarities between threat actors, we also consider the neighboring nodes of the malware authored by those actors (characterized by the \textsf{hasAuthor} relationship). This aggregates information from all the malware authored by the same threat actor for similarity calculation. Again, we use the Jaccard index to compute the similarity between the connected nodes. For example, when looking for the most similar threat actors to \textsf{APT15} from the knowledge graph, we found \textsf{GREF}, \textsf{Boyusec}, and \textsf{Ke3chang}. Among these, \textsf{GREF} and \textsf{Ke3chang} have been listed as being associated with APT15 on the MITRE website \cite{apt15}. This suggests that \textsf{APT15} has been reported under different names in different CTI reports. The other APT group, \textsf{Boyusec}, shares some similarities with APT15, including developing Android surveillance ware, targeting the same messaging apps like Telegram and locations like Uyghur.

\section{Related work}
%\textcolor{blue}{Add rCATT for comparison - results are not encouraging; they use MITRE attack as ground truth, which Tanvir found is inaccurate in Mitre. The attack pattern is linked to files incorrectly.}
Our research closely relates to different areas that support threat intelligence: attack pattern extraction on CTI, information extraction from corpus written in natural language, and cyber threat intelligence knowledge graph. 

\textbf{TTP Extraction from CTI: }Prior research has primarily centered around building rules or models for searching and analyzing IoCs. The limitation of this approach is that there is little to no overlap in the shared information, and no long-term analysis is possible using this intelligence. Most shared intelligence has been limited to tracking known threat indicators such as IP addresses, domain names, and file hashes \cite{liao2016acing,milajerdi2019poirot, shu2018threat}. Some studies \cite{alsaheel2021atlas,shen2019attack2vec,bouwman2020different} use internal enterprise logs to capture threat intelligence and generate attack patterns, but this approach is not scalable. Therefore, none of these can be directly compared to our framework for inference-generating threat intelligence knowledge graphs. In \cite{thein2020paragraph}, Thein et al. propose a neural network approach for classifying sentences of a document into five phases of the cyber kill chain, which represent broad categories for all the phases of a cyberattack. Similarly, TTPDrill~\cite{husari2017ttpdrill} combines dependency parser and heuristics to extract threat actions from a document and maps them to kill chain phases. Luo et al. \cite{luo2021framework} formulate the event extraction as a sequence tagging task and use a bidirectional LSTM network to learn it. AttackKG \cite{li2022attackg} designs technique templates in a graph structure for each attack pattern and maps them to individual MITRE ATT\&CK IDs using a graph alignment algorithm. Our work stands apart from previous research because we develop a novel machine-learning-based algorithm for TTP extraction from unstructured CTI reports. Since our approach does not rely on predefined static patterns, it can discover new attack pattern descriptions from threat reports that are yet to be included in the MITRE standard, as shown in Section \ref{usecase-ttp}. Our approach also results in a lower number of false positives when compared to relevant works.

% \textbf{Discovering and Extracting Threat concepts: } Some approaches \cite{bridges2013automatic} implement a maximum entropy model for automatically labeling cybersecurity concepts. However, more recent studies have developed NER models that adopt a hybrid approach combining multiple methods for NER \cite{yi2020cybersecurity}. Kim et al. \cite{kim2020automatic} built a NER system using a deep bidirectional LSTM-CRF network trained on a combination of features. The work in \cite{dasgupta2020comparative} provides a comparative study between different neural networks, including CNN, LSTM, BERT, and CRF, for cybersecurity NER. OpenCTI reports come from diverse sources; therefore, ambiguity and duplication in threat concepts are research-worthy problems. Word embeddings have been used for entity disambiguation in \cite{vashishth2018cesi} by clustering entities based on their similarity in the embedding space.

\textbf{Named Entity Extraction: }The work in \cite{bridges2013automatic} provides an open-source dataset for cybersecurity named entity extraction. The dataset contains labels for five categories: Version, Application, OS, Vendor, and Relevant. However, the labels were generated automatically using expert rules and were not always accurate. The authors provide three annotated datasets created from different sources -- NVD, MS-Bulletin \cite{msbulletin}, and Metasploit \cite{metasploit}. They reported the annotation F1-score by randomly sampling and hand-annotating a subset of the dataset. The F1-score for the three data sources were 87.5\%, 77.8\%, and 69.1\%, respectively. A more recent dataset published in \cite{kim2020automatic} contains four classes of interest -- URL, Hash, IP address, and Malware. This dataset was manually annotated; however, it lacks the variety of entities in our ontology. The original work used CNN architecture with LSTM and CRF layers and reported an F1-score of 75.1 \%. More recent work on this dataset used transformer-based architecture \cite{Sarhan2021OpenCyKGAO} and achieved an F1-score of 79.8\%. Another study in \cite{Piplai2020CreatingCK} extracted entities from malware after action reports (AAR) with an average F1-score of 77\% for 11 different classes. Recent named entity recognition (NER) models have been built to adopt a hybrid approach combining multiple methods for NER \cite{yi2020cybersecurity}. Kim et al. \cite{kim2020automatic} developed a NER system using a deep bidirectional LSTM-CRF network trained on a combination of features. The work in \cite{dasgupta2020comparative} compares different neural networks, including CNN, LSTM, BERT, and CRF, for cybersecurity NER. 

\textbf{Relation Extraction:} To the best of our knowledge, there is no open-source dataset for cybersecurity relation extraction. Early work in \cite{jones2015towards} used semi-supervised learning with a bootstrapping algorithm for extracting the relation between security entities. They achieved an average F1-score of 82\% on a dataset containing eight different relation types. The work in \cite{pingle2019relext} used a word embedding model for relation extraction in cybersecurity texts. They considered six different types of relations -- hasProduct, hasVulnerability, uses, indicates, mitigates, related-to, and achieved an average F1-score of 92\%. Another study in \cite{guo2021cyberrel} introduced a relation extraction model using the pre-trained BERT model and bidirectional GRU and CRF and achieved an average F1-score of 80.98\%. Recent work on open information extraction \cite{Sarhan2021OpenCyKGAO}, i.e., where the set of relations is not predetermined, achieved an average F1-score of 59.4\%. 

\textbf{Threat Intelligence Knowledge Graph: }Knowledge graph construction in cybersecurity is limited compared to other domains, such as biomedical studies~\cite{nicholson2020constructing}. The study in \cite{Narayanan2018EarlyDO} implements a collaborative framework with the help of semantically rich knowledge representation for the early detection of cybersecurity threats. This system assimilates ontologically defined concepts from multiple sources, such as security bulletins, CVEs, and blogs, and then represents it as a knowledge graph (KG) connected by these concepts. A cybersecurity KG is constructed from open-access CTI in \cite{Piplai2020CreatingCK}, which contains an analysis of various cyber-attacks and is prepared from investigating attacks. This system consists of a custom NER and an entity fusion technique to merge concepts extracted from multiple reports. SEPSES \cite{Kiesling2019TheSK} is a cybersecurity KG populated from multiple heterogeneous cybersecurity data sources and frequently updated. An Extraction, Transformation, and Loading (ETL) periodically checks and updates the KG as new security information becomes available. EXTRACTOR \cite{satvat2021extractor} uses a similar multi-step approach to ours in the process of creating an attack graph from threat reports for threat hunting from kernel logs. However, EXTRACTOR does not include an attack pattern extraction component in the pipeline. EXTRACTOR mainly includes IoCs as nodes in the knowledge graph, which differs from our goal of capturing tactical threat intelligence for malware behavior analysis.
\section{Conclusion}
We propose \textsf{LADDER}, a framework designed to infer attack patterns along with other threat intelligence for existing and emerging threats. We discuss the challenges in extracting threat information (IoCs and attack patterns) from CTI and identify transformer models that work well on cyber threat datasets. \textsf{LADDER} enables security analysts to proactively gain insights into potential ways an emerging threat, described in a CTI report, can impact their internal enterprise network. \textsf{LADDER} also infers attack patterns, which, combined with other classes, {\textsf{Location, Application, OS}}, provides strong evidence for a security analyst when confronting an emerging threat. For future work, we plan to enhance the capabilities of the \textsf{LADDER} framework, integrate a temporal analysis component to track the evolution of attack patterns and techniques over time, evaluate and benchmark the framework's performance against large CTI corpus, and improve integration with system logs providing analysts with a unified interface for threat intelligence.

% \section{Acknowledgments}
% We thank ...

\bibliographystyle{ACM-Reference-Format}
\bibliography{bibliography}

\clearpage
\section*{Appendix}\label{appendix}
\subsection*{Threat Intelligence Concepts}
\label{appendix-concepts}
\textsf{1. Malware} is a central concept of the threat intelligence framework. It is malware reported and analyzed in an open-access CTI, like \textsf{Cerberus}. All other concepts are related to it through a set of relationships. For example, \textsf{Cerberus} (class: Malware) is discovered in \textsf{July, 2019} (class: Time). It \textsf{"tampers with device functionality and steals banking credentials"} (class: Attack Pattern). It targets banking users in \textsf{Spain} (class: Location).
\par 
\textsf{2. Attack Pattern} captures the procedures with which malware performs an attack. For example, an adversary may encrypt files stored on the device to prevent the user from accessing them, with the intent of only unlocking access to the files after a ransom is paid. An open-access CTI report consists of many attack patterns explaining the behavior and procedures of an attack. %MITRE\footnote{\url{https://attack.mitre.org/}} provides a set of commonly used techniques for mobile and enterprise platforms. For the mobile platforms, there are 92 unique techniques. Each attack pattern of malware can be mapped to a standard MITRE ATT\&CK technique. 
Example: Cerberus \textit{steals victim's bank-account credentials} (attack pattern).

\par 
\textsf{3. Malware Type} includes the broad family of malware based on their attack pattern or delivery method. These include, but are not limited to, banking malware, ransomware, adware, spyware, bot, or trojan. Example: \textit{Cerberus} (malware) is a \textit{banking trojan} (MalwareType).
\par 
\textsf{4. Application} includes any software product targeted by malware. Applications can be social media applications or specific businesses like banking apps, e-wallets, games, and utility applications. For example: \textit{Cerberus} (malware) targets a \textit{banking apps} (Application).

%\begin{table}[]
%\centering
%\caption{A table explaining the concepts used in the threat intelligence knowledge graph}
%%\label{table:concepts}
%\resizebox{0.5\textwidth}{!}{%
%\begin{tabular}{|l|l|}
%\hline
%\textbf{Concept} & \textbf{Description}                                                  \\ \hline
%Malware          & Mobile malware reported in the open-access CTI.                               \\ \hline
%Indicator of Compromise (IOC) & Anything that indicates the presence of a malware. \\ \hline
%Attack pattern   & The way of performing an attack by the malware.                       \\ \hline
%Application      & Targeted applications of a mobile malware.                            \\ \hline
%Operating System & Targeted operating system by a mobile malware.                        \\ \hline
%Organization     & Organization/company targeted by the malware.                         \\ \hline
%Person           & Individual who was involved in discovering the malware.     \\ \hline
%Time             & Time to convey when something related to malware occurred. \\ \hline
%Location             & Geographical location(country or city) targeted by the malware. \\ \hline

%\end{tabular}%
%}
%\end{table}

\textsf{5. Operating system} captures the type of OS and kernel targeted by an instance of the malware. Most common OS like Windows, Linux, Mac, Android, and iOS fall in this category. For example: \textit{Cerberus} (malware) targets \textit{Android} (OS).

\textsf{6. Organization} includes a public or private company targeted by a threat attack. Some entities like Facebook, Google, or Twitter can be identified as applications and organizations. In such cases, we follow the context in which the concept appears in the text. For example: \textit{Cerberus} (malware) targets \textit{Google} (Application) users. \textit{Google} (Organization) has updated its security in the play-store to remove harmful applications.

\textsf{7. Person} is an individual who discovered or analyzed the threat attack. Generally, individuals working in a security company are captured by this class.

\textsf{8. Time} conveys when a particular event related to malware occurred. For example: In \textit{June 2019} (Time), ThreatFabric \textit{Organization} found a new \textit{Android} (OS)  malware, dubbed \textit{Cerberus} (malware).

%This information is used to perform longitudinal analysis of malware propagation and evolution.

\textsf{9. Threat Actor} is a person or organization acting with malicious intent either with the development or distribution of malware. Such names are specifically mentioned in the report. For example: \textit{Lazarus} is a group associated with \textit{North Korea} (Location) known for \textit{ransomware}(Malware Type) and attacking banks.

\textsf{10. Location} captures the geographical region, country, or city targeted by a threat attack. Example: \textit{Cerberus} (malware) targeted banking users in \textit{Italy}(Location).

\textsf{11. Indicators of Compromise (IOC)} include anything indicating malware's presence and attack pattern. These include email, hashes, IP addresses, file names, domain names, networks, ports, protocols, and URLs. Example: \textit{corona-apps.apk} (IOC) indicates the distribution of Cerberus malware.

% \textsf{\textbf{Vulnerability}} is a specific flaw in the software or hardware of end-user device exploited by threat actors to perform malicious attacks. A vulnerability can be identified with a standard CVEID if its well-known and has been assigned a name. For example:
% \newline
In addition to these concepts, we establish relationships between different instances using different relationships:
\begin{enumerate}

    % \item variantOf: This relationship captures the variant relationship between malware samples when one malware sample is a variant or modified form of another malware. Example: ERMAC is a \textit{variantOf} Cerberus.
    
    % \item hasAlias: A malware can be identified with a different name. hasAlias captures this relation. This is the only transitive relation, i.e., the head and tail entities are interchangeable. Example: ``Malware Marcher \textit{hasAlias} ExoBot'' is equivalent to ``ExoBot \textit{hasAlias} Marcher''.
    
    \item \textsf{isA}: This relationship classifies a specific malware to a broader family. Example: Cerberus \textit{isA} banking Trojan. 
    
     \item \textsf{targets}: This relationship is between malware and its target, like \textsf{Malware} and \textsf{Location, Organization}, or \textsf{Application}. Example: Cerberus \textit{targets} banking users in Spain.

    \item \textsf{uses}: This relationship is between malware and any entity that it uses to perform an attack, like between \textsf{Malware} and \textsf{Application}, or  \textsf{Malware} and \textsf{AttackPattern}. Example: Cerberus \textit{uses} overlay attacks.

       \item hasAuthor: This relationship connects malware to a threat actor who was responsible for developing the malware. Example: Cerberus \textit{hasAuthor} Kilobyte in the dark web.
 
 \item hasAlias: A malware or threat actor can be identified with a different name. hasAlias captures this relation. This is the only transitive relation, i.e., the head and tail entities are interchangeable. Example: ``Malware Marcher \textit{hasAlias} ExoBot'' is equivalent to ``ExoBot \textit{hasAlias} Marcher''.
 
    \item \textsf{indicates}: This relationship connects an \textit{indicator} to the malware that it represents. Example: Package \textit{com.uxlgtsvfdc.zipvwntdy} \textit{indicates} Cerberus.
    
    \item \textsf{discoveredIn}: This relationship connects malware and the time it was discovered in. Example: Cerberus was \textit{discoveredIn} July 2019. 
    
    \item exploits: This relationship connects a \textsf{Malware} and vulnerability (CVE-IDs) it exploited to perform the attack. Example: Cerberus \textit{exploits} XSS vulnerabilities.
    
    \item \textsf{variantOf}: This relationship connects a \textsf{Malware} to another \textsf{Malware} if one is a variant of another. Example: ERMAC is a \textit{variantOf} Cerberus.
    
       \item \textsf{has}: This is a broad relationship to capture any connection not explained by other relationships, like between \textsf{Malware} and any other entities.
       
%    \item communicatesWith: This relationship connects two entities that communicate with each other over a network. It does not include person to person communication. This relationship can exist between Malware, Organization, Person, ThreatActor and AttackPattern, Malware, Application.
    
\end{enumerate}

\clearpage

\subsection*{Web crawler}
See Algorithm \ref{algo:scraper}.

\begin{algorithm}
\caption{High-Performance Web Crawler}
   \begin{algorithmic}
       \REQUIRE seed\_url
       \ENSURE web\_text, relevant\_urls
       \STATE scraped\_urls  $\leftarrow$ s%crape(seed\_url)
       \STATE check for keywords in scraped\_urls
	 \IF{keyword in url}
	 \STATE url\_queue.add(url)
	 \STATE web\_text.add(text(url))
	 %\STATE relevant\_urls.add(url)
	 \ENDIF
       \FOR{N in generations}
	 \FOR{url in url\_queue}
       \STATE Thread $\leftarrow$ scrape(url)
	 \STATE scraped\_urls  $\leftarrow$ Thread
       \STATE check for keywords in scraped\_urls
	 \IF{keyword in url and url not in url\_queue }
	 \STATE url\_queue.add(url)
	 \STATE web\_text.add(text(url))
	 \STATE relevant\_urls.add(url)
	 \ENDIF
       \ENDFOR
       \ENDFOR
    \end{algorithmic}
\label{algo:scraper}
\end{algorithm}

\subsection*{Regular expressions used to extract IoCs}
\begin{table}[!htbp]
\tiny
\centering
\resizebox{0.8\linewidth}{!}{%
\begin{tabular}{@{}ll@{}}
\toprule
Entity Types & Regular Expression                                                                                                                                           \\ \midrule
FilePath     & {}r'{[}a-zA-Z{]}:\textbackslash{}\textbackslash{}({[}0-9a-zA-Z{]}+)', r'(\textbackslash{}/{[}\textasciicircum{}\textbackslash{}s\textbackslash{}n{]}+)+'{} \\
Email        & {}r'{[}a-z{]}{[}\_a-z0-9-.{]}+@{[}a-z0-9-{]}+{[}a-z{]}+'{}                                                                                                                                                                      \\
SHA256       & {}r'{[}a-f0-9{]}\{64\}|{[}A-F0-9{]}\{64\}'{}                                                                               \\
SHA1         & {}r'{[}a-f0-9{]}\{40\}|{[}A-F0-9{]}\{40\}'{}                                                                                                                                         \\
CVE          & {}r'CVE—{[}0-9{]}\{4\}—{[}0-9{]}\{4,6\}'{}                                                                                                                 \\
IPv4         & {}r'\textasciicircum{}((25{[}0-5{]}|(2{[}0-4{]}|1\textbackslash{}d|{[}1-9{]}|)\textbackslash{}d)(\textbackslash{}.(?!$)|$))\{4\}\$'{}                \\\bottomrule     
\end{tabular}%
}
\caption{Regular expressions used to extract IoCs.}
\end{table}

\subsection*{Attack Patterns for trojan horse, Cerberus}
See Table \ref{tab:MITREattackCerberus-Long} for a complete list of MITRE attack techniques.

\begin{table*}[]
\tiny
\caption{Attack Patterns for \textsf{Cerberus}, extracted from CTI sources, mapped to MITRE ATT\&CK Framework, and ranked in order of occurrence.}
\label{tab:MITREattackCerberus-Long}
\resizebox{\textwidth}{!}{%
\begin{tabular}{llll}
\hline
\textbf{MITRE ID} & \textbf{Name}& \textbf{Description of adversary behavior}& \textbf{Kill-chain Phase}\\ \hline
T1461 &	Lockscreen Bypass &	Bypass device lock-screen &	1: Initial Access\\ \hline
T1404 &	Exploitation for Privilege Escalation &	Exploit vulnerabilities for elevating privileges &	3: Persistence\\ \hline
T1626 &	Abuse Elevation Control Mechanism &	Gain higher-level permissions by taking advantage of built-in control mechanisms. &	4: Privilege Escalation\\ \hline
T1406 &	Obsfucated Files or Information & 	Encrypt, encode or obfuscate the contents of payload or file &	5: Defense Evasion\\ \hline
T1407 &	Download New Code at Runtime &	Download and execute code not included in the original package. &	5: Defense Evasion\\ \hline
T1628 &	Hide Artifacts &	Hide adversary artifacts to evade detection &	5: Defense Evasion\\ \hline
T1629 &	Impair Defense &	Hinder or disable defensive mechanisms of a device &	5: Defense Evasion\\ \hline
T1630 &	Indicator Removal on Host &	Delete, hide, or alter generate adversary artifacts on a device. &	5: Defense Evasion\\ \hline
T1516 &	Input Injection & 	Mimic user interaction abusing Android’s accessibility APIs &	5: Defense Evasion, 12: Impact\\ \hline
T1635 &	Adversary-in-the-Middle &	Position itself between two or more networked devices. &	6: Credential Access\\ \hline
T1417 &	Input Capture &	Capture user input to obtain credentials or information &	6: Credential Access, 9: Collection\\ \hline
T1517 &	Access Notifications &	Collect notifications sent by OS or applications in a mobile device &	6: Credential Access, 9: Collection\\ \hline
T1577 &	Compromise Application Executable &	Modify applications installed on a mobile device &	6: Credential Access, 9: Collection\\ \hline
T1430 &	Location tracking &	Track a device’s physical location &	7: Discovery, 9: Collection\\ \hline
T1428 &	Exploitation of Remote Services &	Exploit remote services of servers, workstations or other resources to gain unauthorized access. &	8: Lateral Movement\\ \hline
T1429 &	Audio Capture & 	Capture audio of a mobile device &	9: Collection\\ \hline
T1512 &	Video Capture &	Video or image files may be written to disk and exfiltrated later. & 9: Collection\\ \hline
T1513 &	Screen capture &	Capture screen to collect additional information about a device. &	9: Collection\\ \hline
T1636 &	Protected user data &	Collect data from permission-backed data stores on a device &	9: Collection\\ \hline
T1481 &	Web Service & 	Use an existing, legitimate web service for transferring data to and from a device &	10: Command \& Control\\ \hline
T1639 &	Exfiltration Over Alternative Protocol &	Steal data by exfiltrating it over different protocol than the existing C\&C. &	11: Exfiltration\\ \hline
T1471 &	Data Encrypted for Impact &	Encrypt files on a device to prevent user access &	12: Impact\\ \hline
T1582 &	SMS Control &	Delete, alter or send SMS messages. &	12: Impact\\ \hline
T1616 &	Call Control &	Make, forward or block phone calls &	12: Impact, 9: Collection\\ \hline
\end{tabular}%
}
\end{table*}

\end{document}